\documentclass[onecolumn]{revtex4}
\usepackage{amsmath}
\usepackage{amsfonts,color}
\usepackage{amssymb,float}
\usepackage{mathtools}
\usepackage[utf8]{inputenc}
\usepackage{url}
\usepackage{graphicx}
\usepackage{refstyle}
\usepackage{wasysym}
\usepackage{accents}
\usepackage{graphicx}
\usepackage{color}
\usepackage[dvipsnames]{xcolor}
\usepackage{hyperref}
\usepackage{anysize}
\usepackage[labelfont=bf]{caption}
\usepackage[nodisplayskipstretch]{setspace}
\usepackage[ margin=2cm]{geometry}
\usepackage{amsmath,latexsym}
\usepackage{amsfonts}
\usepackage{amssymb}
\usepackage{amsthm}
\usepackage{caption}
\usepackage{pdfpages}
\usepackage{booktabs}
\usepackage{adjustbox}
\usepackage{fontenc}
\usepackage{graphicx}
\usepackage{dcolumn}
\usepackage{bm}
\usepackage{hyperref}
\usepackage{listings}
\usepackage{verbatim}   
\usepackage{multirow}
\usepackage{subfigure}
\usepackage{float}
\usepackage{enumerate}
\usepackage{cancel}
\usepackage{tensor}

\hypersetup{
    colorlinks=true,
    linkcolor=blue,
    filecolor=magenta,      
    citecolor=red
}

\begin{document}

\title{Modified Yang-Lee theory for nonlocal gravitational potential and their phase transition}

\author{Carlos Ar\'aoz Alvarado}
\email{carlosaraoz@ciencias.unam.mx}
\affiliation{Instituto de Ciencias Nucleares, Universidad Nacional Aut\'{o}noma de M\'{e}xico, 
Circuito Exterior C.U., A.P. 70-543, M\'exico D.F. 04510, M\'{e}xico.}

\author{Celia Escamilla-Rivera}
\email{celia.escamilla@nucleares.unam.mx}
\affiliation{Instituto de Ciencias Nucleares, Universidad Nacional Aut\'{o}noma de M\'{e}xico, 
Circuito Exterior C.U., A.P. 70-543, M\'exico D.F. 04510, M\'{e}xico.}


\begin{abstract}
In this paper, we describe the extension to study the thermodynamics of the structure formation in the large scale Universe in the nonlocal gravity formalism using standard statistical mechanics. From the derivation of the grand partition function in a modified version of the Yang-Lee theory, we obtained the corresponding thermodynamics properties that can be in consistency with a Bose-Einstein dark matter framework and derive its gravitational phase transition.
\end{abstract}

\maketitle

\section{Introduction}

The effects of modifications to the Newtonian (interaction) potential energy in the thermodynamics of clustering of galaxies can be studied in the formalism of Statistical Mechanics \cite{Khanday:2021kjy,hameeda2021large}. It has been considered that approximating galaxies as point-like particles under a gravitational partition function, we can  deduce certain thermodynamic quantities of the system at hand, from which we can investigate the structure formation at large scales \cite{Upadhyay:2018ykz}.

Following these ideas, in Capozziello \emph{et al}\cite{Capo} was presented a study where we can get information of the effects in the properties of galaxies and clusters of galaxies by the consideration of a logarithmic correction under the nonlocal gravity formalism to the Newtonian (interaction) potential energy between particles, due to the inspection of the thermodynamics of such systems. In particular, they assume a grand canonical ensemble frame and derive the grand canonical partition function. Under these assumptions it is possible to deduce thermodynamic quantities, specifically, Helmholtz free energy F, the Entropy S, the internal energy U, the pressure P, and the chemical potential $\mu$, which can set the dependence of the Helmholtz free energy F and the internal energy U on the same parameters.

Under such considerations, nonlocal theories of General Relativity (GR) have recently been studied \cite{Modesto:2021soh,Hameeda:2021zsu,Bittencourt:2020lgu}. 
A nonlocal modification of the Einstein-Hilbert action has been studied by Deser and Woodard\cite{PhysRevLett.99.111301}, 
\begin{equation}
   \mathcal{S}=\frac{1}{2\kappa^2}\int d^4x\sqrt{-g}[R(1+f(\square^{-1}R))],
   \label{eqn:actionnon}
\end{equation}
where $g=det(g_{\mu \nu})$, $R$ is the Ricci scalar and $f(\square^{-1}R)$ is a function called \textit{distortion function} of the nonlocal term $\square^{-1}R$, observe that if we settle $f(\square^{-1}R)=0$ in  Eq.(\ref{eqn:actionnon}) we recover the Einstein-Hilbert action. This nonlocal term is given by the retarder Green's function denote by:
\begin{equation}
    \mathcal{G}[f](x)=(\square^{-1}f)(x)=\int d^4x'\sqrt{-g(x')}f(x')G(x,x').
\end{equation}
 The nonlocality is established by the inverse of the d'Alembert operator as we can see from the first term at the left of the latter equation. A local representation of Eq.(\ref{eqn:actionnon}) can be obtained by the introduction of an auxiliary scalar field such that it satisfies $\square^{-1}R=\phi$, therefore $\square \phi=R$. 
The weak-field limit approximation produces corrections to the Newtonian interaction potential. These corrections can appear as polynomials or logarithmic forms as discussed in Dialektopoulos \emph{et al}\cite{Dia}.

In these scenarios we can found that nonlocal GR simulates dark matter as manifestation of the nonlocality of the gravitational interaction. In this line of study, nonlocal scenarios called Yang-Lee describes some types of phase transitions that could give the distribution of zeros of the canonical partition function in the fugacity plane.
Therefore, it is possible to adopt a grand canonical formalism for a system with a large number of galaxies and obtain the canonical partition function in terms of the fugacity and consequently search for real zeros of the partition function. In order to find a proper expression, the gravitational partition function with effects caused by the cosmological constant has been used to study the effect of the expansion of the universe on the clustering of galaxies \cite{Ham}.
Other examples in this topic have been discussed by Saslaw \emph{et al}\cite{Saslaw}, where the partition function has been used to study the gravitational many-body clustering of galaxies in an expanding Universe\footnote{This landscape resembles a form of phase transition using a Newtonian (interaction) potential energy background.}. With this function,  we can obtain the thermodynamics of the system (specifically the chemical potential) and then use Yang-Lee phase transition theory to analyze the phase transition of this system using the zeros of the complex fugacity $z$.
 
Using these idea, in this paper we study the thermodynamics of a specific system with a large number of galaxies and a logarithmic correction to the Newtonian gravity interaction potential through a nonlocal gravitational partition function and, in the landscape of Yang-Lee theory, we found the conditions for a phase transition of this system using a grand canonical ensemble framework. 
 
The paper is divided as follows: 
in Sec.~\ref{sec:partition} we define the nonlocal gravitational partition function for a system with a large number of galaxies and their logarithmic corrections to the Newtonian potential energy.
In Sec.~\ref{sec:Thermodynamics} we use the nonlocal gravitational partition function to obtain the thermodynamic quantities and study their evolution.
In Sec.~\ref{sec:MYang-Lee} we compute the partition function in terms of its roots, then we derive different thermodynamic quantities and study their behavior. 
Sec.~\ref{sec:Yang-Lee} explore the idea of a phase transition in the frame of Yang-Lee theory. 
In Sec.~\ref{sec:Specific Heat} we discuss the consequences of the evolution of specific heat with respect to a critical temperature and found the relationship with the effective potential obtained from reducing the Klein-Gordon equation to a Schr\"{o}dinger-like equation 
in a Schwarzschild spacetime configuration. 
Finally, in Sec.~\ref{sec:conclusions} we present our final results.


\section{ Nonlocal Gravitational Partition Function}
\label{sec:partition}

From this point forward, we are going to approximate galaxies as point-like particles\cite{Capo}. We consider a system with a large number of galaxies and embrace an ensemble of cells, with the same volume V (or radius $R_1$) and with an average density $\rho$. Since the cells have a variable number of galaxies and their total energy can vary between them, the system can be analyzed in the framework of a grand canonical ensemble. We can consider this approximation because the distance between galaxies is much larger than their proper size. 

In this scenario, a system of galaxies interacting with a nonlocal gravitational potential can be described by the following nonlocal gravitational partition function

\begin{equation}
    Z(T,V)=\frac{1}{\lambda^{3N}N!}\int d^{3N}pd^{3N}r\ \exp\left(-\left[\sum_{i=1}^N\frac{p_i^2}{2M}+\Phi_{nl}(r)\right]T^{-1}\right),
\end{equation}
where $N$ is the number of galaxies, $M$ is the mass, $p_i$ are the moments of several galaxies, $N!$ takes into account the distinguis-hability of classical particles, $\lambda$ 
normalizes the phase space volume cell and it is known as a nonlocality parameter, $T$ is the average temperature and $\Phi_{nl}(r)$ is the nonlocal gravitational potential energy. 

Rewriting the gravitational partition function: 
\begin{eqnarray}
        Z(T,V) 
 &=&\frac{1}{\lambda^{3N}N!}\int d^{3N}p\ \exp\left(-\sum_{i=1}^N\frac{p_i^2}{2M}T^{-1}\right)\int d^{3N}r \exp\left(-\Phi_{nl}(r)T^{-1}\right),
     \label{eqn:zinit}
    \end{eqnarray}
we can derive    
\begin{equation}
   \int d^{3N}p\exp\left(-\sum_{i=1}^N\frac{p_i^2}{2M}T^{-1}\right)= \int d^{3N}p\exp\left(-\frac{1}{2M}T^{-1}\sum_{i=1}^Np_i^2\right),
   \end{equation}
   where
\begin{equation}
    \sum_{i=1}^Np_i^2=p_1^2+p_2^2+...+p_N^2=p\cdot p=(p_1,...,p_N)\cdot (p_1,...,p_N),
\end{equation}
therefore
\begin{equation}
  \int d^{3N}p\exp\left(-\sum_{i=1}^N\frac{p_i^2}{2MT}\right)= \int d^{3N}p\exp\left(-\frac{1}{2MT}p\cdot p\right).
   \end{equation}
Performing the integration over momentum space we obtain
\begin{equation}
 \int d^{3N}p\exp\left(-\frac{1}{2MT}p\cdot p\right)=(2M\pi T)^{\frac{3N}{2}}.
  \label{eqn:momspaint}
   \end{equation}
In general, since the nonlocal and local terms represent central forces the nonlocal gravitational potential energy $\Phi_{nl}(r_1, r_2, . . . , r_N )$ depends on the relative position vector $r_{ij} = |r_i- r_j |$,
where
   \begin{equation}
   \begin{aligned}
      \int d^{3N}r \exp\left(-\Phi_{nl}(r)T^{-1}\right) 
            &= \int\cdot \cdot\cdot\int  \prod_{1\leq i < j \leq N} \exp\left(-\frac{\Phi_{nl}(r_{ij})}{T}\right)d^{3N}r,
      \end{aligned}
\end{equation}
we can write the total nonlocal potential energy $\Phi_{nl}(r_1, r_2, . . . , r_N )$ as the sum of the nonlocal potential energies as
\begin{equation}
    \Phi_{nl}(r_1, r_2, . . . , r_N )=\sum_{1\leq i < j \leq N}\Phi_{nl}(r_{ij})=\sum_{1\leq i < j \leq N}\Phi_{nl}(r).
\end{equation}
By defining the configuration integral $Q_N(T,V)$ since it involves an integral over all configurations, or positions, of the point-like particles as:
  \begin{equation}
      Q_N(T,V)=\int\cdot \cdot\cdot\int  \prod_{1\leq i < j \leq N} \exp\left(-\frac{\Phi_{nl}(r_{ij})}{T}\right)d^{3N}r,
      \label{eqn:confint}
  \end{equation}
we can substitute Eq.(\ref{eqn:confint}) 
 and Eq.~(\ref{eqn:momspaint}) in Eq.~(\ref{eqn:zinit}) to obtain the following gravitational partition function 
 \begin{equation}
     Z_N(T,V)=\frac{1}{N!}\left(\frac{2\pi M T}{\lambda^2}\right)^{\frac{3N}{2}}Q_N(T,V).
 \end{equation}
Writing a nonlocal potential energy for galaxies as
 \begin{equation}
     \left(\Phi_{i,j}\right)_{nl}=-\frac{GM^2}{r_{ij}}+\frac{GM^2}{\lambda}\ln\left(\frac{r_{ij}}{\lambda}\right),
 \end{equation}
where the former corresponds to the Newtonian (interaction) potential energy between particles and the latter corresponds to the logarithmic correction induced by the nonlocal gravity.  
Since galaxies have an extended structure since  are not actually point-like particles,
we need to add to the nonlocal potential energy a softening parameter $\epsilon$. Then, we obtain
\begin{equation}
     \left(\Phi_{ij}\right)_{nl}=-\frac{GM^2}{(r_{ij}^2+\epsilon ^2)^\frac{1}{2}}+\frac{GM^2}{\lambda}\ln\left(\frac{(r_{ij}^2+\epsilon ^2)^\frac{1}{2}}{\lambda}\right).
     \label{eqn:potwe}
\end{equation}
 We can use a Mayer function \cite{Schr}, that is a nonlocal two-point interaction function for galaxies in Eq.~(\ref{eqn:potwe}) defined by:
 \begin{equation}
     f_{ij}=e^{-\frac{(\Phi_{ij})_{nl}}{T}}-1.
     \label{eqn:2pifij}
     \end{equation}
This equation is equal to zero in the absence of interactions (like in an ideal gas),  and it is non-zero for galaxies that interact. As
\begin{equation}
   f_{ij}+1=e^{-\frac{(\Phi_{ij})_{nl}}{T}}=e^{-\frac{\Phi_{nl}(r_{ij})}{T}},
   \label{eqn:2pint}
   \end{equation}
   we obtain
 \begin{equation}
  \prod_{1\leq i<j \leq N}\exp{-\frac{\Phi_{nl}(r_{ij})}{T}}= \prod_{1\leq i < j \leq N}(f_{ij}+1)=(f_{12}+1)(f_{13}+1)...(f_{N-1,N}+1).
\end{equation}
Substituting the latter in Eq.~(\ref{eqn:confint}) we obtain
\begin{equation}
    Q_N(T,V)=\int\cdot \cdot\cdot\int\left[(f_{12}+1)(f_{13}+1)\cdot \cdot\cdot(f_{N-1,N}+1)\right]d^3r_1 \cdot \cdot\cdot d^3r_N.
    \label{eqn:qn}
\end{equation}
   Using Eq.~(\ref{eqn:2pint}) we notice that
   \begin{equation}
    -T\ln(f_{ij}+1)=(\Phi_{ij})_{nl},
       \label{eqn:desp1}
   \end{equation}
   theferore, from Eq.~(\ref{eqn:potwe}) and Eq.~(\ref{eqn:desp1}) we obtain
     \begin{equation}
  f_{ij}=\left(\frac{(r_{ij}^2+\epsilon ^2)^\frac{1}{2}}{\lambda}\right)^{-\frac{GM^2}{\lambda T}}\exp\left(\frac{GM^2}{(r_{ij}^2+\epsilon ^2)^\frac{1}{2}T}\right)-1.
   \end{equation}   
Expanding as a power series we have
   \begin{equation}
   f_{ij}=-1+\left(\frac{(r_{ij}^2+\epsilon ^2)^\frac{1}{2}}{\lambda}\right)^{-\frac{GM^2}{\lambda T}}\sum_{n=0}^{\infty}\frac{1}{n!}\left( \frac{GM^2}{T}\right)^n\frac{1}{(r_{ij}^2+\epsilon ^2)^{\frac{n}{2}}}.
   \label{eqn:fij}
     \end{equation} 
   Defining $l=\frac{GM^2}{T\lambda}$ from Eq.~(\ref{eqn:qn}) and Eq.~(\ref{eqn:fij}) by considering $N=2,i=1,j=2$ and considering the calculations presented in \ref{app:def} we obtain
 \begin{eqnarray}
        Q_2(T,V) &=&\int\int1-1+\left(\frac{(r_{12}^2+\epsilon ^2)^\frac{1}{2}}{\lambda}\right)^{-l}\sum_{n=0}^{\infty}\frac{1}{n!}\left( \frac{GM^2}{T}\right)^n\frac{1}{(r_{12}^2+\epsilon ^2)^{\frac{n}{2}}} d^3r_1d^3r_2
        \nonumber \\ &&
        =\int\int\lambda^l\sum_{n=0}^{\infty}\frac{1}{n!}\left( \frac{GM^2}{T}\right)^n      {}_2F_1\left(\frac{n+l}{2},b;b;-\left(\frac{r_{12}}{\epsilon}\right)^2\right)\frac{1}{\epsilon^{n+l}}d^3r_1d^3r_2. \quad\quad
        \label{eqn:hyper}
\end{eqnarray}  
We define the following function   
   \begin{equation}
       x:=\frac{3}{2}(Gm^2)^3\rho T^{-3}=\beta \rho T^{-3},
       \label{eqn:x}
   \end{equation}
where $\beta=\frac{3}{2}(GM^2)^3$ to get $[x]=N^3 m^3 kg K^{-3}$.
Using the dilute approximation, where all pairs of particles will be far enough apart that 
   $(\Phi_{i,j})_{nl}/T\ll 1$, and therefore the factor $e^{-\frac{(\Phi_{ij})_{nl}}{T}}$ is near to unity. 
   We have
\begin{equation}
    -\ln(f_{ij}+1)=\frac{(\Phi_{i,j})_{nl}}{T},
\end{equation}
    \begin{equation}
     -T\ln(f_{ij}+1)=-\frac{GM^2}{(r_{ij}^2+\epsilon ^2)^\frac{1}{2}}+\frac{GM^2}{\lambda}\ln\left[\frac{(r_{ij}^2+\epsilon ^2)^\frac{1}{2}}{\lambda}\right],
   \end{equation}

 \begin{equation}
 f_{ij}+1 =\exp\left( \frac{GM^2}{(r_{ij}^2+\epsilon ^2)^\frac{1}{2}T}-\frac{GM^2}{\lambda T}\ln\left[(r_{ij}^2+\epsilon ^2)^\frac{1}{2}\right]+\frac{GM^2}{\lambda T}\ln\lambda \right).
 \end{equation}
 Taking the first order expansion we obtain
   \begin{equation}
 f_{ij} = \frac{GM^2}{(r_{ij}^2+\epsilon ^2)^\frac{1}{2}T}-\frac{GM^2}{\lambda T}\ln\left[(r_{ij}^2+\epsilon ^2)^\frac{1}{2}\right]+\frac{GM^2}{\lambda T}\ln\lambda,
 \end{equation}  
 with
 \begin{equation}
    Q_N(T,V)=\int \cdot \cdot \cdot \int \prod_{1\leq i<j \leq N}\left(1+\frac{GM^2}{(r_{ij}^2+\epsilon ^2)^\frac{1}{2}T}-\frac{GM^2}{\lambda T}ln\left((r_{ij}^2+\epsilon ^2)^\frac{1}{2}\right)+\frac{GM^2}{\lambda T}\ln\lambda  \right)d^{3N}r.
 \end{equation}
 Finally, we obtain
 \begin{equation}
     Q_N(T,V)=V^N(1+\alpha x)^{N-1},
 \end{equation}
where
 \begin{eqnarray}
       \alpha&=& \frac{2R}{3\lambda}\left[\ln \left(\frac{\lambda}{R}\right)+\frac{3\lambda}{2R}\sqrt{\frac{\epsilon^2}{R^2}+1} +\frac{\epsilon^3}{R^3}\tan^{-1}\frac{R}{\epsilon}-\frac{3\lambda \epsilon^2}{2R^3}\ln\left(\frac{\sqrt{\frac{\epsilon^2}{R^2}+1}+1}{\frac{\epsilon}{R}} \right)   \right] \nonumber \\ &&
      - \frac{2R}{3\lambda}\left(-\frac{1}{2}\ln\left(\frac{\epsilon^2}{R^2}+1 \right)-\frac{\epsilon^2}{R^2}+\frac{1}{3}   \right).
      \label{eqn:alpha}
 \end{eqnarray}
  Defining $\gamma=\frac{\epsilon}{R}$ and $\sigma=\frac{\lambda}{R}$ in  Eq.~(\ref{eqn:alpha}) we have 
 \begin{eqnarray}
      \alpha&=& \frac{2}{3\sigma}\ln \left(\sigma \right)+(\gamma^2+1)^{\frac{1}{2}}+\frac{2}{3\sigma}\gamma^3\tan^{-1}(\gamma^{-1})-\gamma^2\ln\left[\frac{(\gamma^2+1)^{\frac{1}{2}}+1}{\gamma} \right]  \nonumber \\
      &&- \frac{1}{3\sigma}\ln(\gamma^2+1)-\frac{2\gamma^2}{3\sigma}+\frac{2}{9\sigma}.
 \end{eqnarray}
 Finally, we can write the gravitational partition function as
 \begin{equation}
     Z_N(T,V)=\frac{1}{N!}\left(\frac{2\pi MT}{\lambda^2} \right)^{\frac{3N}{2}}V^N(1+\alpha x)^{N-1}.
     \label{eqn:cpf}
 \end{equation}
With this gravitational partition function, we can obtain the thermodynamics of the astrophysical system.


\section{Nonlocal Gravity Thermodynamics}
\label{sec:Thermodynamics}

\begin{figure}[t]
    \centering
    \includegraphics[width=0.32\textwidth]{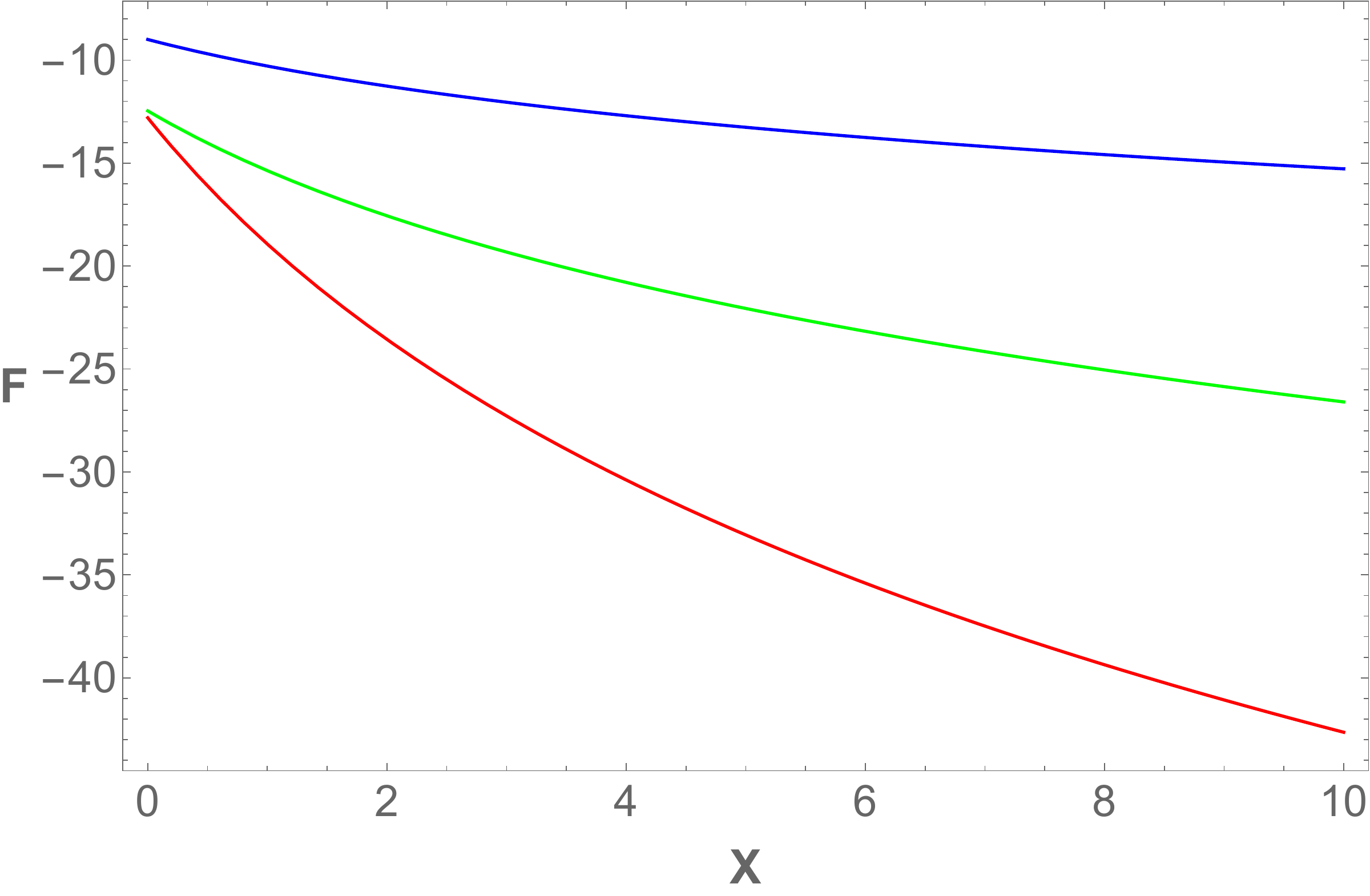}
        \includegraphics[width=0.32\textwidth]{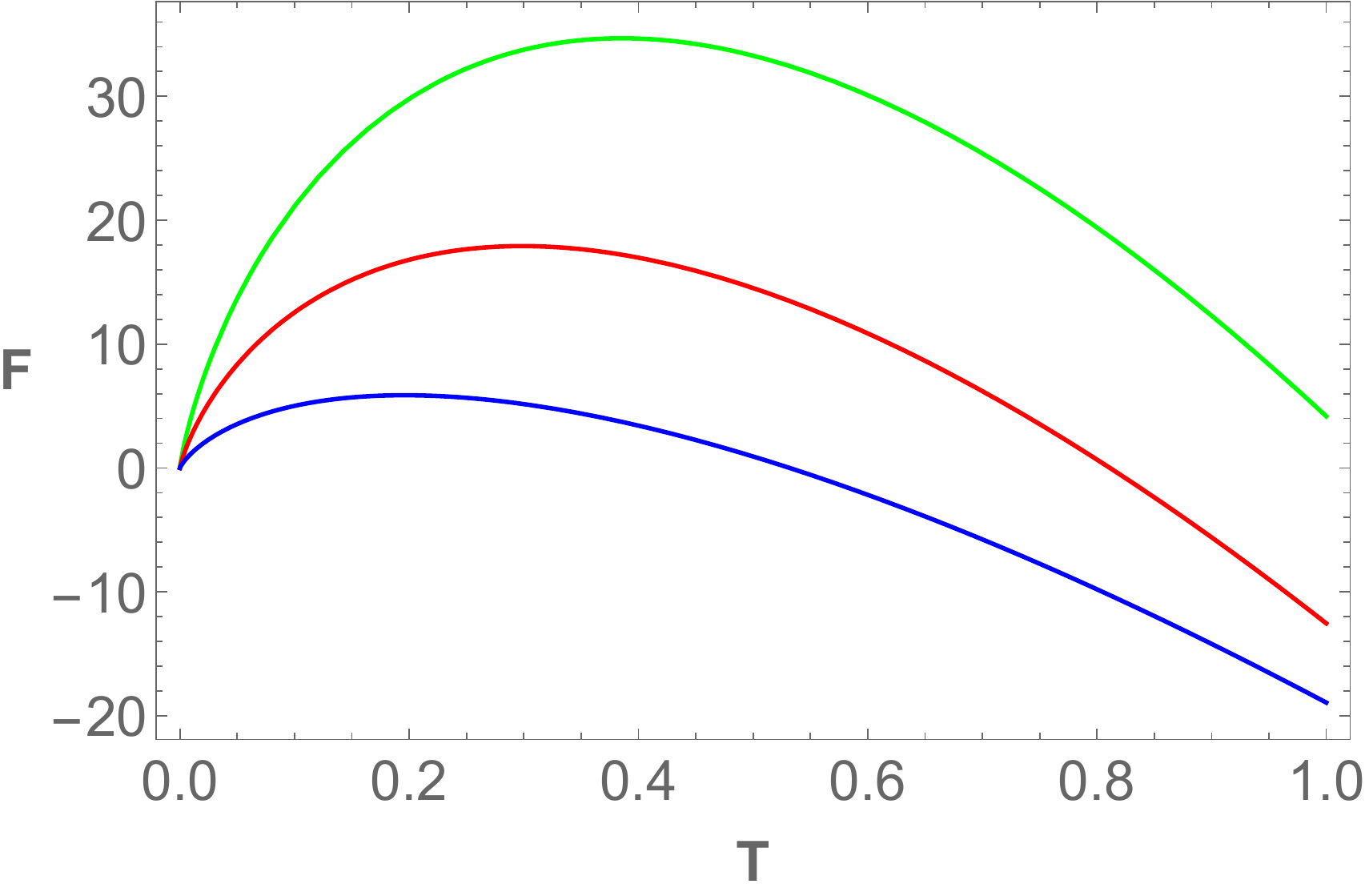}
     \includegraphics[width=0.32\textwidth]{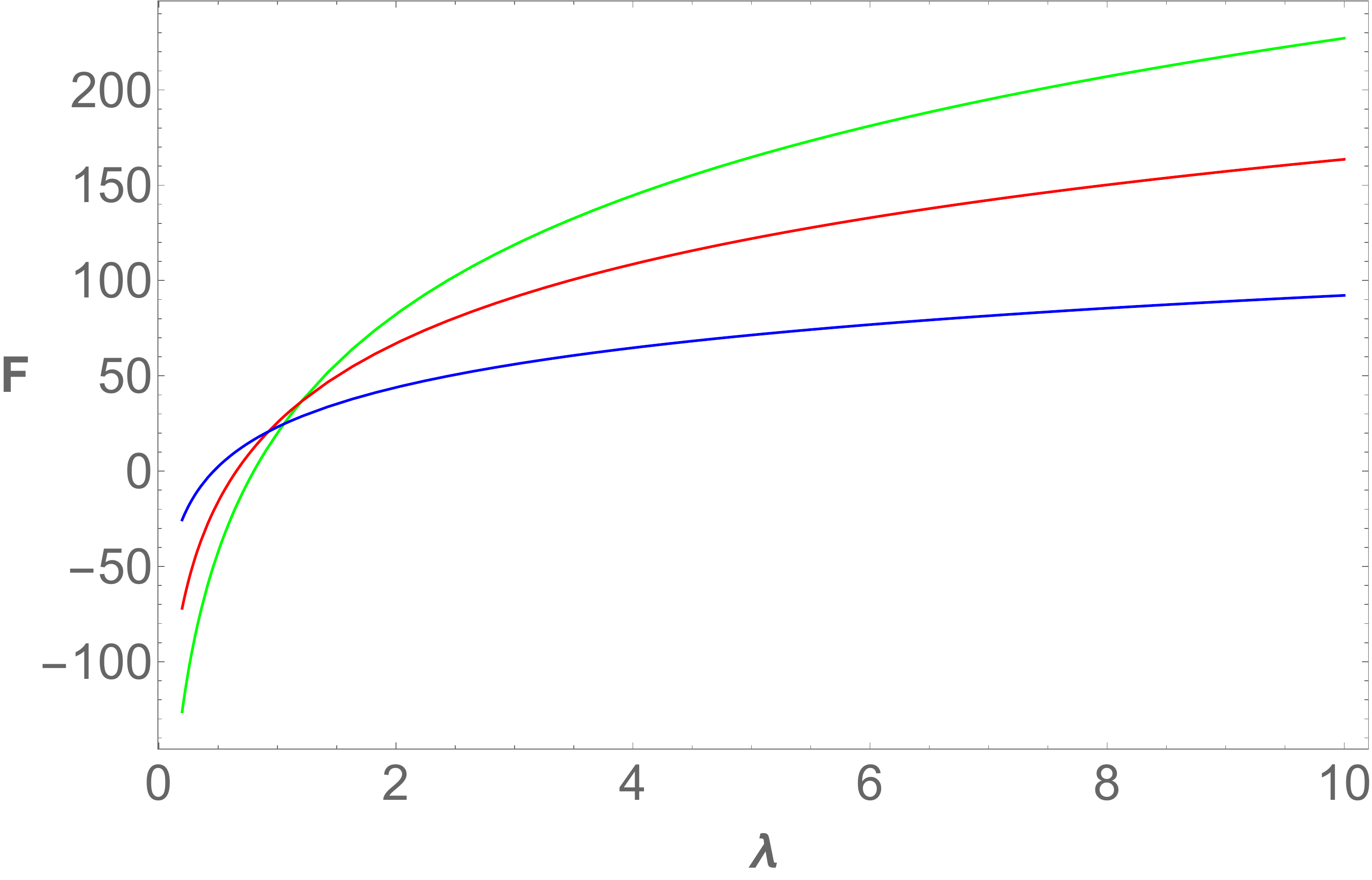}
    \caption{\textit{Left:} Helmholtz free energy dependence on $x$ with: $\lambda=1$, (a) $N=10$  (green); (b) $N=20$  (red);\\ (c) $N=5$  (blue). \textit{Middle:} Helmholtz free energy dependence on $T$ with: $x=1$,   (a) $N=60$ (green); (b) $N=40$ (red); (c)  $N=20$  (blue). \textit{Right:} Helmholtz free energy dependence on $\lambda$ with:  $x = 1$   and $N=50$, (a) $T=0.6$ (green); (b) $T=0.4$ (red);  (c) $T=0.2$  (blue). We have assumed unit values for all the other parameters.}
    \label{fig:Helmholtz}
\end{figure}

By definition, the Helmholtz free energy is $F=-T\ln Z_N(T,V)$, therefore
   \begin{equation}
      F=-T\ln\left[\frac{1}{N!}\left(\frac{2\pi MT}{\lambda^2} \right)^{\frac{3N}{2}}V^N(1+\alpha x)^{N-1}\right].
   \end{equation}
   In  Figure \ref{fig:Helmholtz} we analyze the dependence of the Helmholtz free energy $F$ on several parameters, where we made explicit the dependence of $\alpha$ on $\gamma$ and $\sigma$. In this evolution Fig.~\ref{fig:Helmholtz}a we show the Helmholtz free energy dependence on $x$. We notice that the Helmholtz free energy is a decreasing function of $x$ for $\lambda=1$, without considering beforehand the value of the number of galaxies $N$. Fig.~\ref{fig:Helmholtz}b shows the Helmholtz free energy as a function of $T$. We distinguish that the Helmholtz free energy has a maximum and this point appears to change with the number of galaxies $N$. Fig.~\ref{fig:Helmholtz}c shows the Helmholtz free energy dependence on $\lambda$. We notice that the Helmholtz free energy increases with the value of $\lambda$ nevertheless the rate of this increase depends on the value of the temperature $T$. 
   
\begin{figure}[t]
    \centering
    \includegraphics[width=0.32\textwidth]{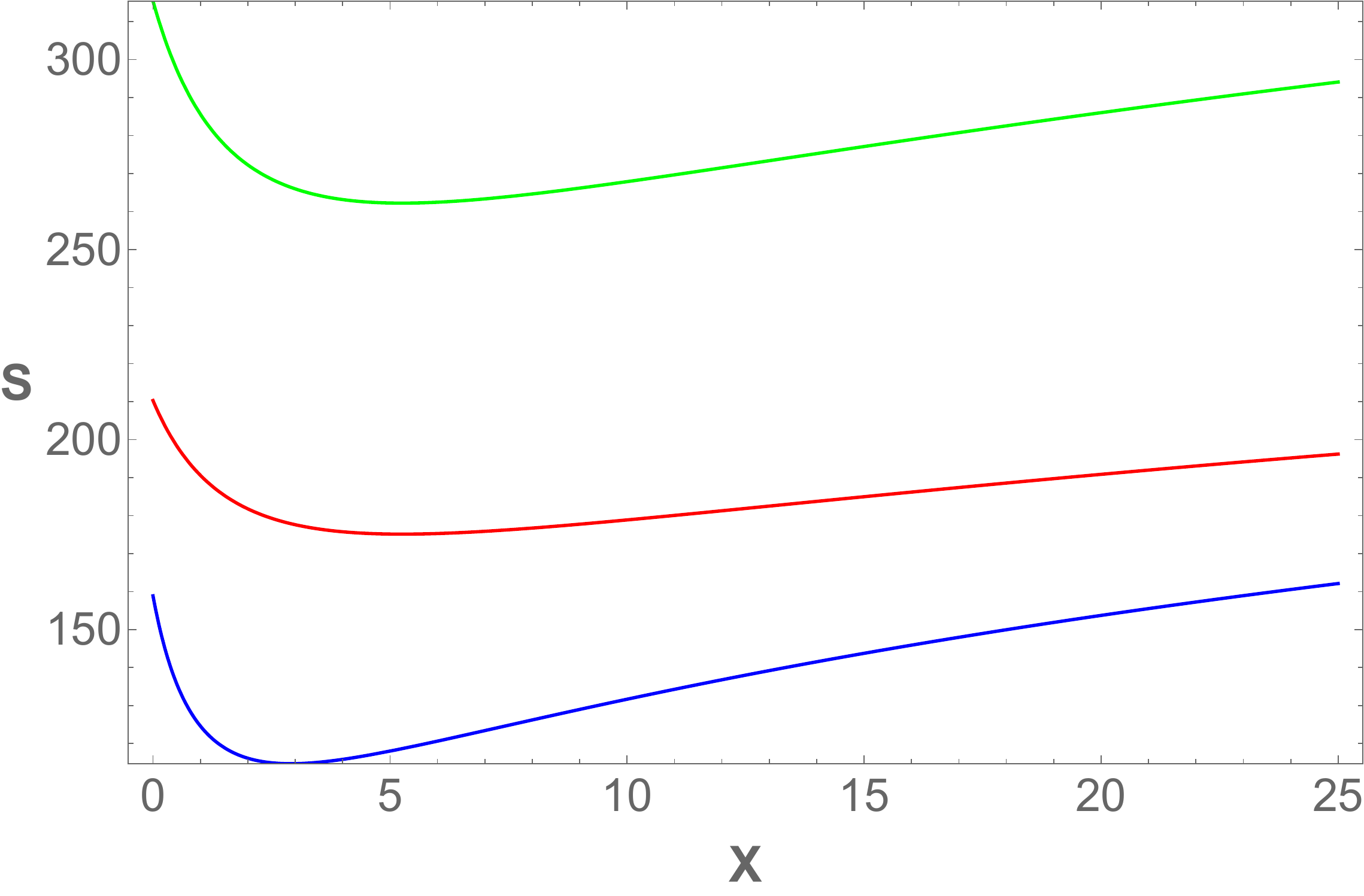}
      \includegraphics[width=0.32\textwidth]{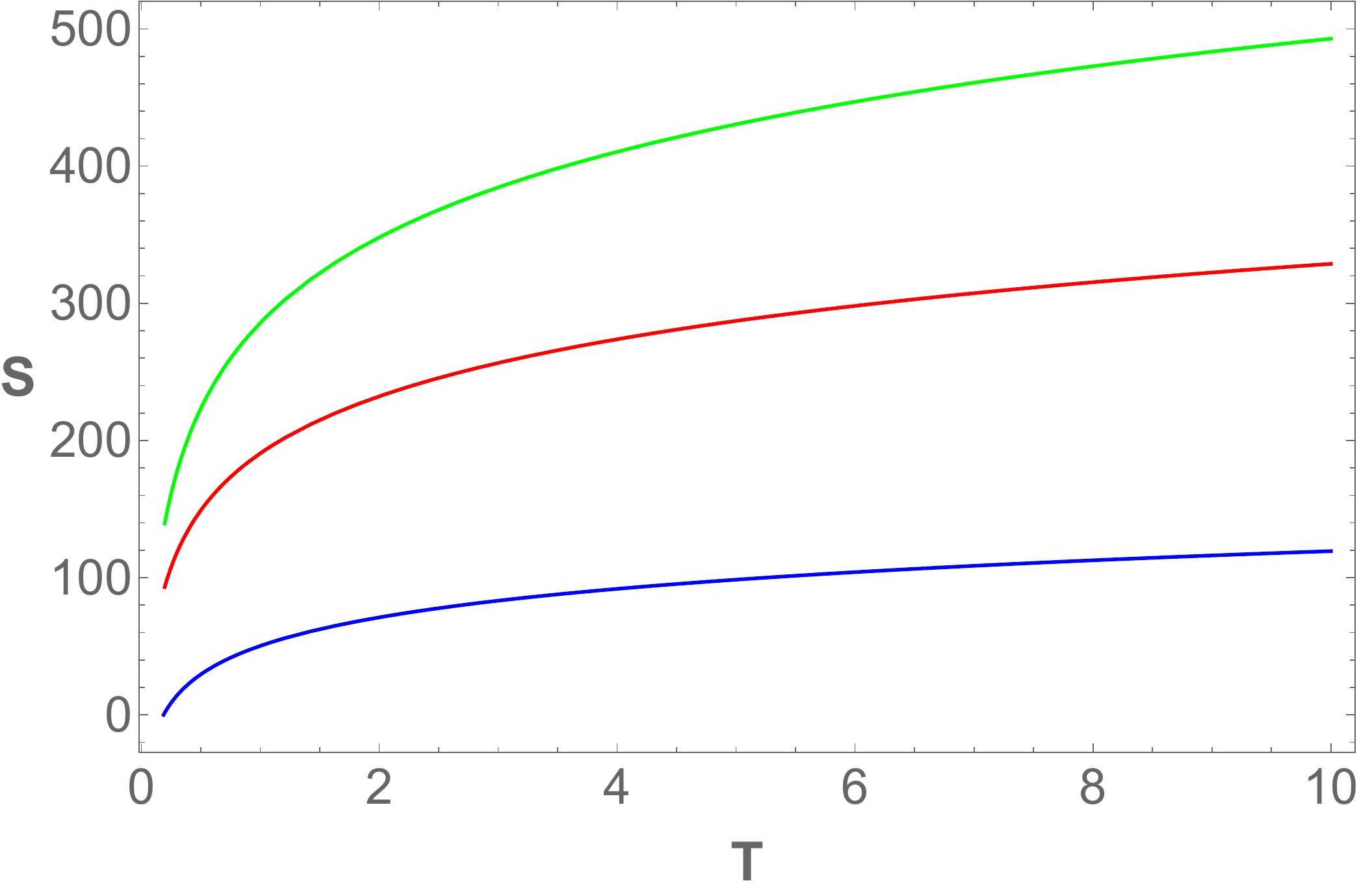}
       \includegraphics[width=0.32\textwidth]{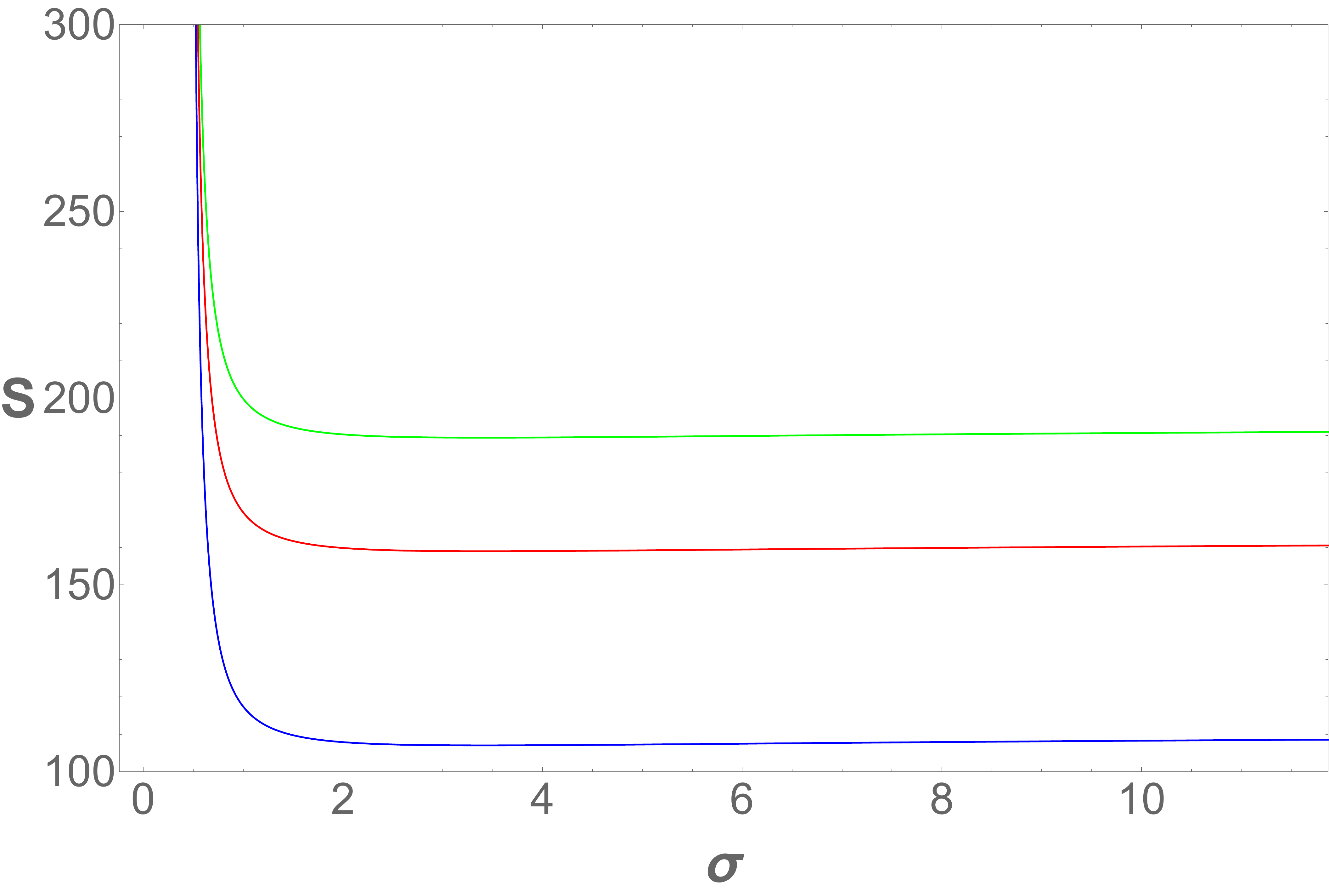}
      \caption{\textit{Left:} Entropy dependence on $x$ with: $\lambda=1$,   (a) $N=10$  (green); (b) $N=20$  (red); (c) $N=5$  (blue). \textit{Middle:} Entropy dependence on $T$ with:  $x=1$, (a) $N=60$ (green); (b) $N=40$ (red); (c)  $N=20$  (blue). \textit{Right:} Entropy dependence on $\sigma$ with:   $x = 1$   and $N=50$,  (a) $T=0.6$ (green); (b) $T=0.4$ (red); (c)  $T=0.2$  (blue).  We have assumed unit values for all the other parameters.}
    \label{fig:Entropy}
\end{figure}

Also, by definition, the entropy $S$ can be obtained from this latter Helmholtz free energy by
 \begin{eqnarray}
      S &=&-\left(\frac{\partial F}{\partial T}\right)_{N,V}\nonumber \\
     &=&\frac{\partial }{\partial T} \left(T\ln\left[\frac{1}{N!}\left(\frac{2\pi MT}{\lambda^2} \right)^{\frac{3N}{2}}V^N(1+\alpha x)^{N-1}\right]\right).
 \end{eqnarray}
   Defining $A=\frac{1}{N!}V^N, B=\frac{2\pi M}{\lambda^2}$ we can write
    \begin{equation}
      S 
     = \frac{3N}{2}\ln B+\frac{3N}{2}+(N-1)\ln(1+\alpha x)-\frac{3(N-1)\alpha x}{(1+\alpha x)}+\ln\left(\frac{1}{N!}V^N\right)+\frac{3N}{2}\ln T.
 \end{equation}
   From the last two terms in the latter equation, we obtain using $\rho^{-1}=\frac{V}{N}$ and Stirling's approximation $ln(N!)\approx Nln(N)-N$ the following
   \begin{equation}
         \ln\left(\frac{1}{N!}V^N\right)+\frac{3N}{2}\ln(T)=  N\ln(\rho^{-1}T^{\frac{3}{2}})+N,
   \end{equation}
   and
      \begin{equation}
    S= \frac{3N}{2}\ln(\frac{2\pi M}{\lambda^2})+\frac{5N}{2}+(N-1)\ln(1+\alpha x)-\frac{3(N-1)\alpha x}{(1+\alpha x)}+ N\ln(\rho^{-1}T^{\frac{3}{2}}).
 \end{equation}
For a large value of $N$ using $N-1\approx N$ we have
  \begin{equation}
      S= \frac{3N}{2}\ln(\frac{2\pi M}{\lambda^2})+\frac{5N}{2}+N\ln(1+\alpha x)-\frac{3N\alpha x}{(1+\alpha x)}+ N\ln(\rho^{-1}T^{\frac{3}{2}}).
 \end{equation}
In the plots presented in Figure \ref{fig:Entropy} we show the dependence of the entropy $S$ on some parameters, with explicit dependence  of $\alpha$ on $\gamma$ and $\sigma$. Fig.~\ref{fig:Entropy}a, shows the dependence of the entropy on $x$. We notice that there is a minimum for the entropy and this point, analogous to the latter case, changes with the number of galaxies $N$. In Fig.~\ref{fig:Entropy}b the entropy is plotted in terms of  $T$. We observe that the entropy increases with the value of $T$ nevertheless, the rate of this increase depends on the value of $N$. Fig.~\ref{fig:Entropy}c illustrates the entropy as a function of  $\sigma$. We notice that the entropy decreases with the value of $\sigma$ but the rate of this decrease depends on the value of the temperature $T$.

 Defining
 \begin{equation}
     S_0=\frac{5N}{2}+\frac{3N}{2}\ln\left(\frac{2\pi M}{\lambda^2}\right),
 \end{equation}
 and
 \begin{equation}
  B_l=\frac{\alpha x}{(1+\alpha x)}, 
  \label{eqn:Bl}
 \end{equation} 
 where
  \begin{equation}
  \frac{S}{N}
     =\ln(\rho^{-1}T^{\frac{3}{2}})+\ln(1+\alpha x)-3B_l+\frac{S_0}{N}.
 \end{equation}
 Using $N-1\approx N$ 
  \begin{equation}
        F=-T\ln\left[\frac{1}{N!}\left(\frac{2\pi MT}{\lambda^2} \right)^{\frac{3N}{2}}V^N(1+\alpha x)^{N}\right].
   \end{equation}
The internal energy $U$ of a system of galaxies can 
be obtained using Stirling's approximation by $U=F+TS$
    \begin{equation}
           U 
          = \frac{3}{2}NT(1-2B_l).
          \label{eqn:ieu}
   \end{equation}

\begin{figure}[t]
    \centering
    \includegraphics[width=0.32\textwidth]{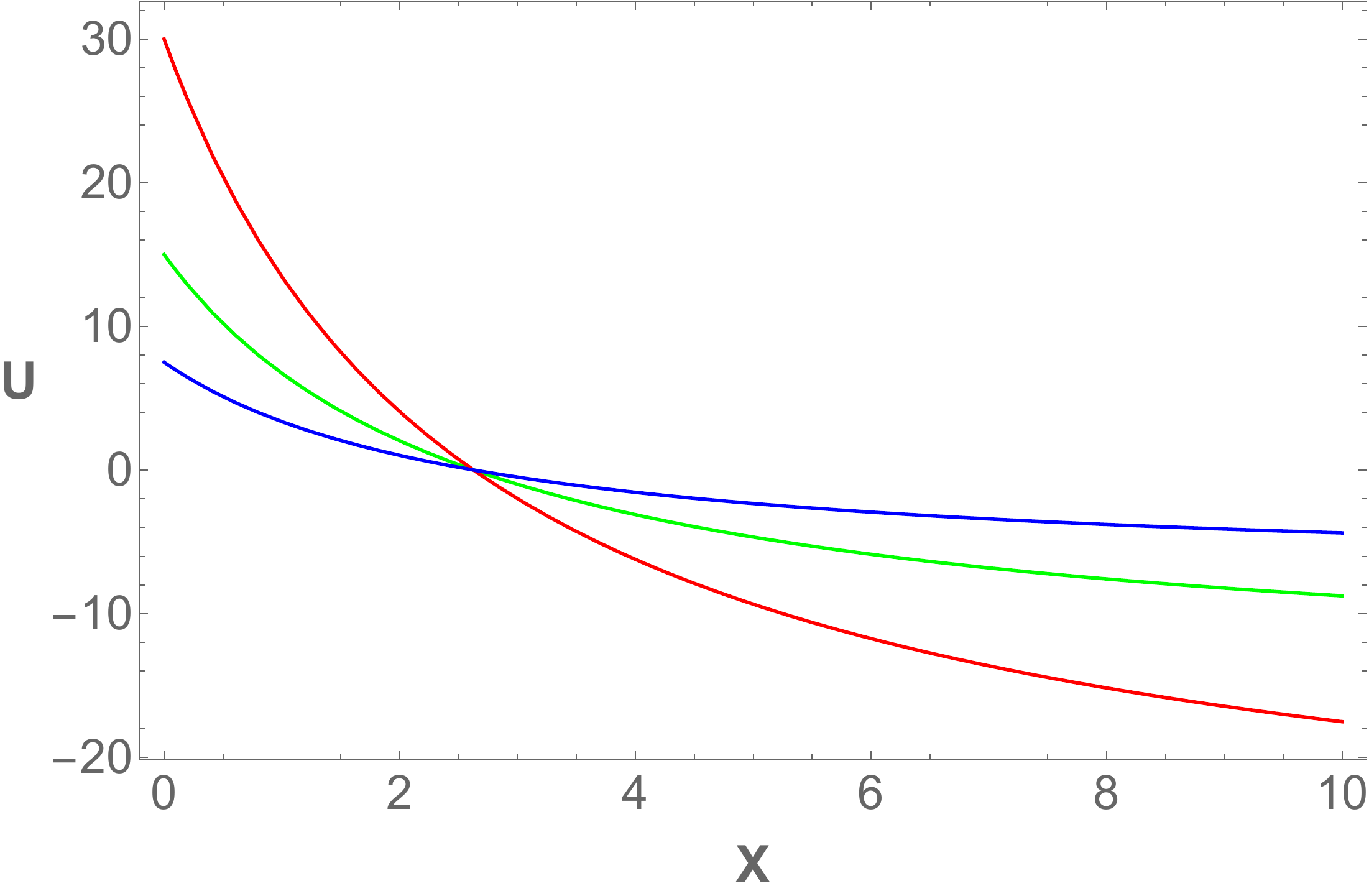}
    \includegraphics[width=0.32\textwidth]{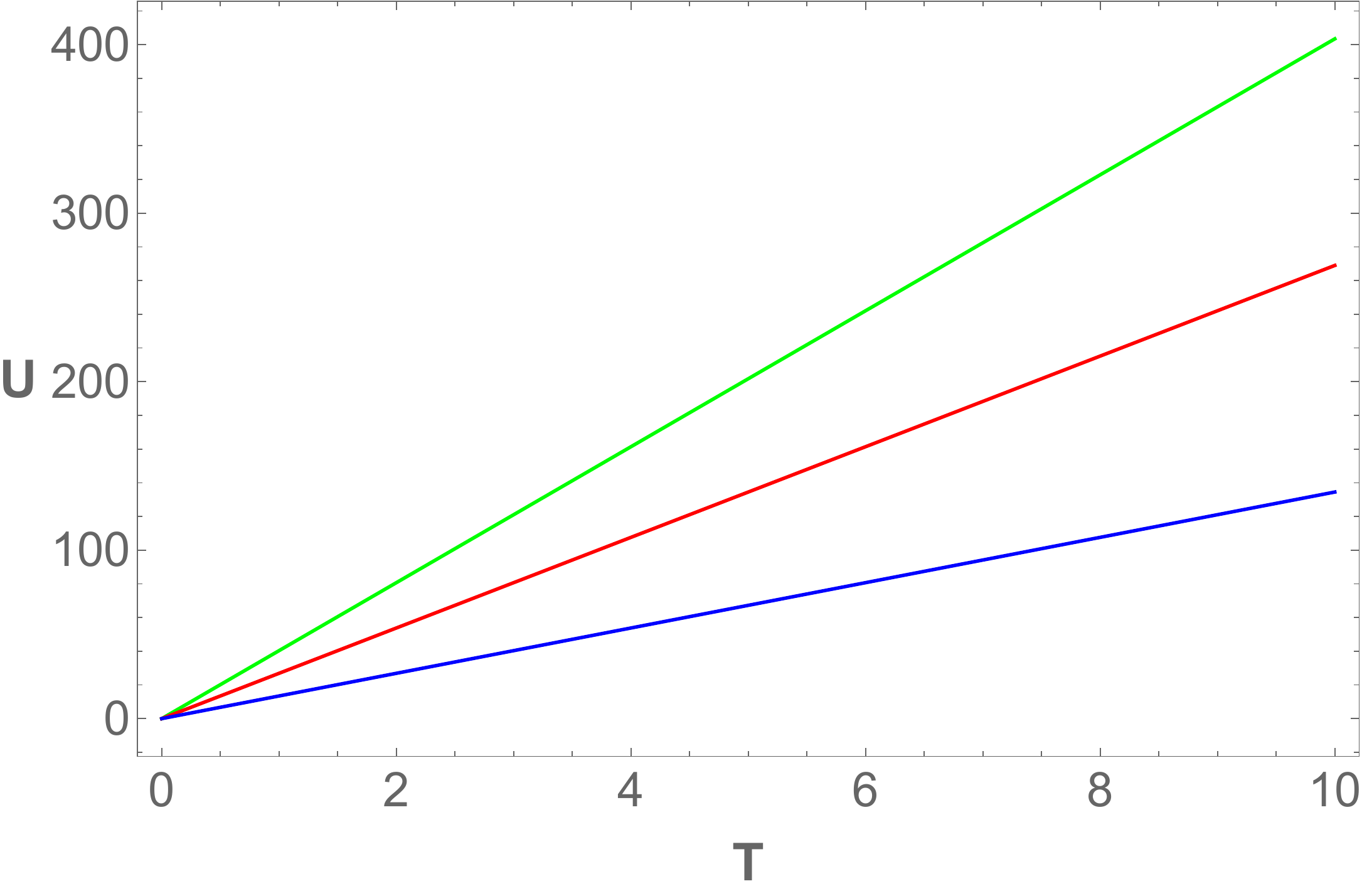}
\includegraphics[width=0.32\textwidth]{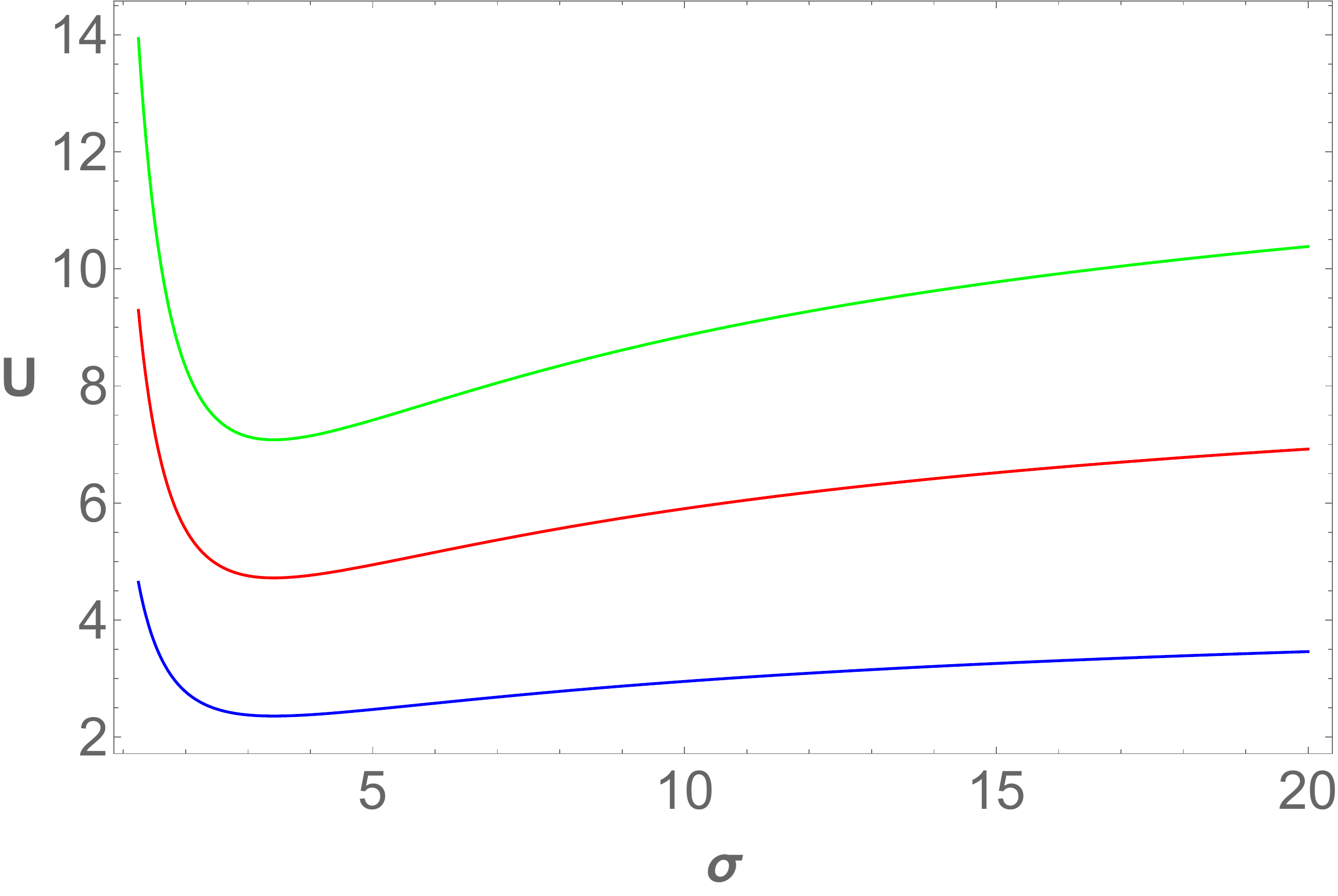}
    \caption{\textit{Left:} Internal energy dependence on $x$ with: $\lambda=1$,    (a) $N=10$  (green); (b) $N=20$  (red); (c) $N=5$  (blue). \textit{Middle:} Internal energy dependence on $T$ with: $x=1$,  (a) $N=60$ (green); (b) $N=40$ (red); (c)  $N=20$  (blue). \textit{Right:} Internal energy dependence on $\sigma$ with:   $x = 1$   and $N=50$, (a) $T=0.6$ (green); (b) $T=0.4$ (red);  (c)  $T=0.2$  (blue). We have assumed unit values for all other parameters.}
    \label{fig:Interenergy}
\end{figure}

We illustrate the internal energy $U$ in Figure \ref{fig:Interenergy}, where we make explicit the dependence of $B_l$ on $\alpha$ and $x$, and also make explicit the dependence of $\alpha$ on $\gamma$ and $\sigma$. Fig.~\ref{fig:Interenergy}a shows the internal energy as a function of $x$, we detect that the internal energy is a decreasing function of $x$ for $\lambda=1$ regardless of which value of the number of galaxies $N$ we take. We investigate in Fig.~\ref{fig:Interenergy}b the behavior of the internal energy on $T$. We see that the internal energy increases with the value of $T$ nevertheless the rate of this increase is dependent on the value of the number of galaxies $N$. The dependence of the internal energy on  $\sigma$ is plotted in Fig.~\ref{fig:Interenergy}c. We can observe that there is a minimum for the internal energy, and this minimum does not seem to change with the value of the temperature $T$.

  The pressure $P$ can be deduced from
  \begin{equation}
          P =-\left(\frac{\partial F}{\partial V}\right)_{N,T}
            =T \frac{\partial }{\partial V} \left[\ln\left(\frac{1}{N!}\left(\frac{2\pi MT}{\lambda^2} \right)^{\frac{3N}{2}}V^N(1+\alpha \beta \frac{N}{V} T^{-3})^{N}\right)\right].
   \end{equation}
   Defining $A=\frac{1}{N!}\left(\frac{2\pi MT}{\lambda^2} \right)^{\frac{3N}{2}}$ and $B=\alpha \beta N T^{-3}$,
    \begin{eqnarray}
          P &=&T \frac{\partial }{\partial V} \left(\ln\left[\frac{1}{N!}\left(\frac{2\pi MT}{\lambda^2} \right)^{\frac{3N}{2}}V^N(1+\alpha \beta \frac{N}{V} T^{-3})^{N}\right]\right)\nonumber \\
                     &=& 
           \frac{TN}{V}\left[]1 -\frac{B}{\left(1+\frac{B}{V} \right)V} \right],
                  \label{eqn:pr}
   \end{eqnarray}  
   for
   \begin{equation}
     1 -\frac{B}{\left(1+\frac{B}{V} \right)V}
       =1 -B_l,
   \end{equation}
   finally
  \begin{equation}
    \therefore ~ P=\frac{NT}{V}(1 -B_l).
    \label{eqn:p}
   \end{equation}
Furthermore, we can obtain the chemical potential $\mu$ using $\rho=\frac{N}{V}$ and Stirling's approximation $ln(N!)\approx N\ln(N)-N$ 
    \begin{eqnarray}
          \mu &=&\left(\frac{\partial F}{\partial N}\right)_{V,T}\nonumber \\
          &=&-T\frac{3}{2}\ln\left(\frac{2\pi M}{\lambda^2} \right)+T\ln(\rho T^{-\frac{3}{2}})-T\ln(1+\alpha x) -TB_l.
           \label{eqn:mu}
   \end{eqnarray}

In a grand canonical ensemble, the probability that $N$ galaxies (particles) occupy a cell is the sum over all of their energy states can be described by
 \begin{equation}
      F(N)=\sum_i\frac{e^{\frac{N\mu}{T}}e^{\frac{-U_n}{T}}}{Z_G(T,V,Z)}
      =\frac{e^{\frac{N\mu}{T}}\sum_i e^{\frac{-U_n}{T}}}{Z_G(T,V,Z)},
 \end{equation}
since
 \begin{equation}
     Z_N(T,V)=\sum_ie^{\frac{-U_n}{T}},
 \end{equation}
   \begin{equation}
    F(N)=\frac{e^{\frac{N\mu}{T}}Z_N(T,V)}{Z_G(T,V,Z)}.
   \label{eqn:fn}
 \end{equation} 
 where $ Z_G(T,V,Z)$ is the grand partition function defined by
 \begin{equation}
     Z_G(T,V,Z)=\sum_{N=0}^\infty z^NZ_N(T,V),
 \end{equation}
where $z=e^{\frac{\mu}{T}}$ is the fugacity.
 
\begin{figure}[t]
    \centering
    \includegraphics[width=0.32\textwidth]{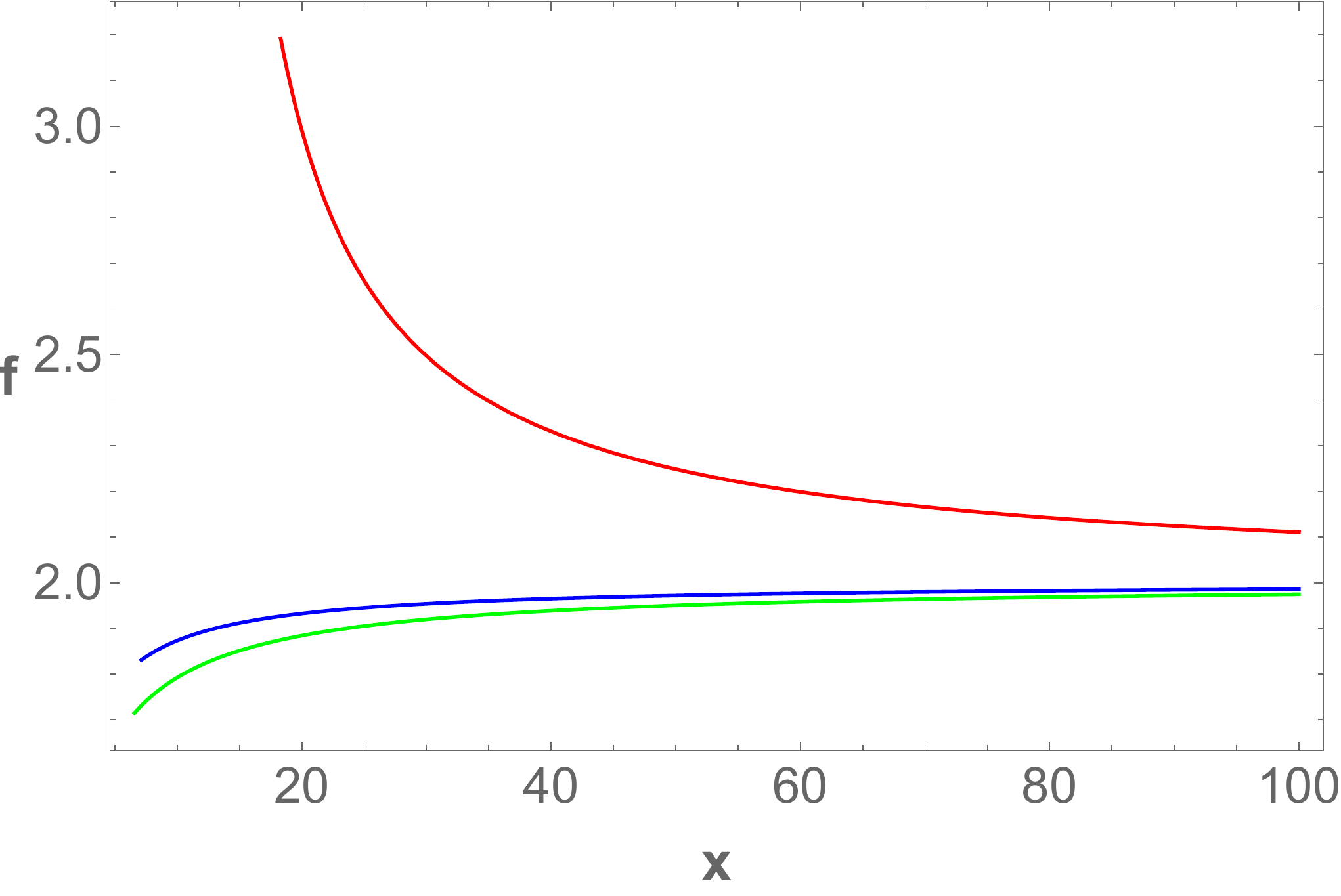}
    \includegraphics[width=0.32\textwidth]{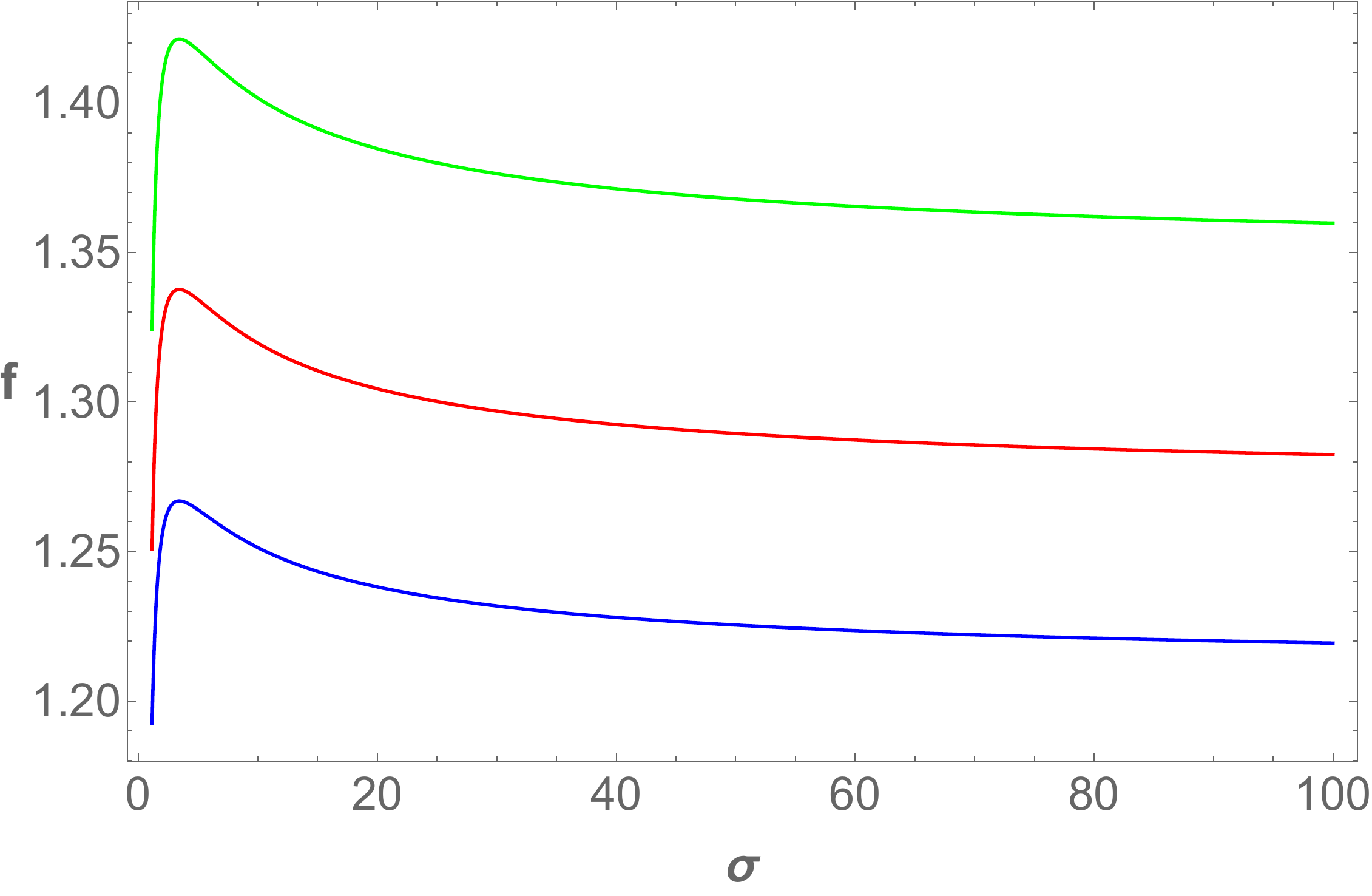}
   \includegraphics[width=0.32\textwidth]{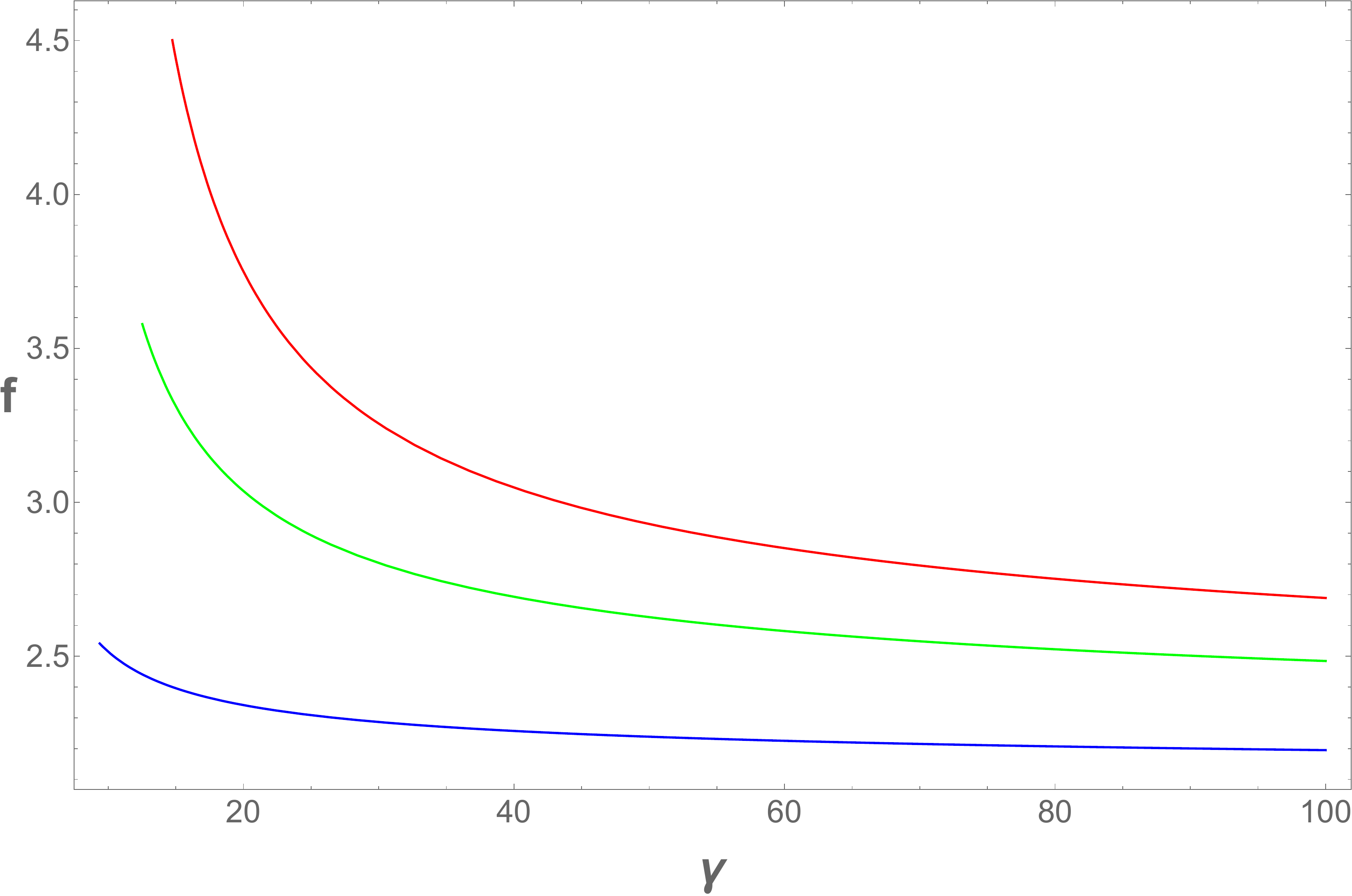}
     \caption{\textit{Left:} Distribution function dependence on $x$ with: $\gamma=1$, (a) $\sigma=1$  (green); (b) $\sigma=2/3$ (red); (c) $\sigma=2$  (blue). \textit{Middle:} Distribution function dependence on $\sigma$ with:  $\gamma= 1$, (a) $x=1$ (green); (b)  $x=0.7$ (red); (c)  $x=0.5$  (blue). \textit{Right:} Distribution function dependence on $\gamma$ with: $\sigma = 1$, (a) $x=1$ (green); (b) $x=0.8$ (red); (c)  $x=2$  (blue). We have assumed unit values for all other parameters.}
    \label{fig:F(N)}
\end{figure}

Due to that the probability has to be normalized $F(N)=1$, where the grand partition function can be thought of as a normalization constant, i.e. is the factor that is needed to get the sum of the probabilities equal to one
 \begin{equation}
     1= F(N)=\frac{e^{\frac{N\mu}{T}}Z_N(T,V)}{Z_G(T,V,Z)},
 \end{equation} 
 \begin{equation}
 Z_G(T,V,Z)=e^{\frac{N\mu}{T}}Z_N(T,V). 
    \label{eqn:Zg}
 \end{equation}    
Considering
   \begin{eqnarray}
   e^{\frac{N\mu}{T}}&=&\exp\left(\frac{N}{T}\left(T\ln(\rho T^{-\frac{3}{2}})-T\ln(1+\alpha x)-T\frac{3}{2}\ln\left(\frac{2\pi M}{\lambda^2} \right) -TB_l\right)\right) \nonumber \\
   &=&(\frac{N}{V} T^{-\frac{3}{2}})^N(1+\alpha x)^{-N}\left(\frac{2\pi M}{\lambda^2} \right)^{-\frac{3N}{2}} \exp(-NB_l), 
 \end{eqnarray}  
notice that
 \begin{equation}
     \left(1+\alpha x\right)^{-N} = 
     \left(1+\frac{B_l}{(1-B_l)}\right)^{-N},
 \end{equation}
     \begin{equation}
 e^{\frac{N\mu}{T}}
     =(\frac{N}{V} T^{-\frac{3}{2}})^N\left(1+\frac{B_l}{(1-B_l)}\right)^{-N}\left(\frac{2\pi M}{\lambda^2} \right)^{-\frac{3N}{2}} \exp(-NB_l).
 \end{equation}  
 Using $N-1\approx N$ in
 \begin{equation}
     Z_N(T,V)=\frac{1}{N!}\left(\frac{2\pi MT}{\lambda^2} \right)^{\frac{3N}{2}}V^N(1+\alpha x)^{N},
 \end{equation}
in Eq.~(\ref{eqn:Zg}) we obtain
 \begin{equation}
 Z_G(T,V,Z)=e^{\frac{N\mu}{T}}Z_N(T,V)
          =\frac{N^N}{N!}e^{-NB_l}.
    \label{eqn:Zgg}
 \end{equation}       
From another form of the Stirling's approximation  
 \begin{equation}
     N!\approx N^Ne^{-N}\sqrt{2\pi N}\approx N^Ne^{-N}, \quad  \Rightarrow \quad \frac{N^N}{N!}\approx e^{N},
 \end{equation}
  and
 \begin{equation}
 Z_G(T,V,Z)=e^{N}e^{-NB_l}=e^{N-NB_l}, \quad \therefore \quad Z_G(T,V,Z)=e^{N(1-B_l)},
 \end{equation}     
with
   \begin{equation}
    \ln(Z_G(T,V,Z))=N(1-B_l). 
 \end{equation}   
 From Eq.~(\ref{eqn:p}) we notice that 
 \begin{equation}
 P\frac{V}{T}=\frac{V}{T}\frac{NT}{V}(1 -B_l)=N(1-B_l)=\ln(Z_G(T,V,Z)).
   \end{equation}
Therefore, Eq.~(\ref{eqn:fn}) can be rewritten as
 \begin{eqnarray}
F(N)&=&\frac{e^{\frac{N\mu}{T}}Z_N(T,V)}{Z_G(T,V,Z)},\nonumber \\
   &=&\frac{N^N}{N!}  e^{-NB_l-N(1-B_l)}(1+\alpha x)^{-1},
 \end{eqnarray}  
and
 \begin{equation}
 (1+\alpha x)^{-1}=\frac{1}{(1+\alpha x)}= \frac{1+\alpha x- \alpha x}{(1+\alpha x)} = 1- \frac{\alpha x}{1+\alpha x}=1-B_l,
 \end{equation}
 where
  \begin{eqnarray}
F(N)  
   &=&\frac{N^N}{N!} e^{-NB_l-N(1-B_l)}1-B_l, \nonumber \\
 &=&\frac{N^N(1-B_l)}{N!} e^{-N}\nonumber \\
   &=&e^{N}(1-B_l) e^{-N}\nonumber \\
   &=&1-B_l.
   \label{eqn:dfn}
 \end{eqnarray}   
In  Figure \ref{fig:F(N)} we present the dependence of the distribution function $F(N)=f$ (where we changed $F(N)$ for $f$ just to simplify the notation of the plots) on various parameters, with explicit dependence of $B_l$ on $\alpha$ and $x$, and also with explicit dependence of $\alpha$ on $\gamma$ and $\sigma$. Fig.~\ref{fig:F(N)}a shows the distribution function as a function of $x$. We can observe that  for small values of $\sigma$ the distribution function is a decreasing function of $x$ and for large values of $\sigma$ an increasing function of $x$, however, it appears that they tend to the same curve. The dependence of the distribution function on  $\sigma$ is represented in Fig.~\ref{fig:F(N)}b. We can detect that the distribution function has a maximum and this maximum seems to change with the value of $x$. Fig.~\ref{fig:F(N)}c exhibits the dependence of the distribution function on  $\gamma$. We recognize that the distribution function is a decreasing function of $\gamma$ but the rate of decrease depends on the value of the temperature $T$.

 \section{Roots of the Partition Function}
 \label{sec:MYang-Lee}
 
Considering the grand canonical description of the system. For a given volume $V$, a specific temperature $T$, with $b$ being the single-particle excluded volume and where $M =\frac{V}{b}$ is the maximum number of particles that  the system can contain, the grand partition function is given by:
\begin{equation}
    \Xi(T,z)=\sum_{N=0}^M Z_N(V,T) z^N.
\end{equation}
As $Z_N(V,T)$ is the canonical partition function with a fixed number of particles $N$ and $=e^{\frac{\mu}{T}}$ is the fugacity (activity) of the system.

 The grand partition function  $\Xi(T,z)$  was rewritten due to the fundamental theorem of algebra in terms of its $M$ roots $z_i = z_i(T)$, ($i=1,\cdot \cdot \cdot ,M$) (\emph{cf}. Bena \emph{et al}. \cite{bena}).
\begin{equation}
   \Xi(T,z)=\zeta \prod_{i=1}^M \left(1-\frac{z}{z_i(T)}\right),
\end{equation}
where $\zeta$ is a multiplicative constant, that will be omitted in the foregoing.
We notice that
\begin{equation}
    \ln\Xi=\ln\left[\prod_{i=1}^M \left(1-\frac{z}{z_i(T)}\right)\right]=\sum_{i=1}^M\ln \left(1-\frac{z}{z_i(T)}\right).
    \label{eqn:lnxi}
\end{equation}
By definition the entropy $S$ is 
\begin{eqnarray}
  S
  &=&\ln\Xi+T\left(\frac{\partial \ln\Xi}{\partial T}\right)_{V,\mu} \nonumber \\
    &=&\sum_{i=1}^M\ln \left(1-\frac{e^{\frac{\mu}{T}}}{z_i(T)}\right)+T\sum_{i=1}^M\left(\frac{e^{\frac{\mu}{T}}}{z_i(T)-e^{\frac{\mu}{T}}}\right)\left(\frac{\mu}{T^2}+\frac{1}{z_i(T)}\frac{\partial  z_i(T)}{\partial T} \right).
   \end{eqnarray}
Finally,
\begin{equation}
S=\sum_{i=1}^M\left[\ln \left(1-\frac{e^{\frac{\mu}{T}}}{z_i(T)}\right)+T\left(\frac{e^{\frac{\mu}{T}}}{z_i(T)-e^{\frac{\mu}{T}}}\right)\left(\frac{\mu}{T^2}+\frac{1}{z_i(T)}\frac{\partial  z_i(T)}{\partial T} \right)\right]. 
    \label{eqn:syl}
\end{equation}

\begin{figure}[t]
    \centering
        \includegraphics[width=0.32\textwidth]{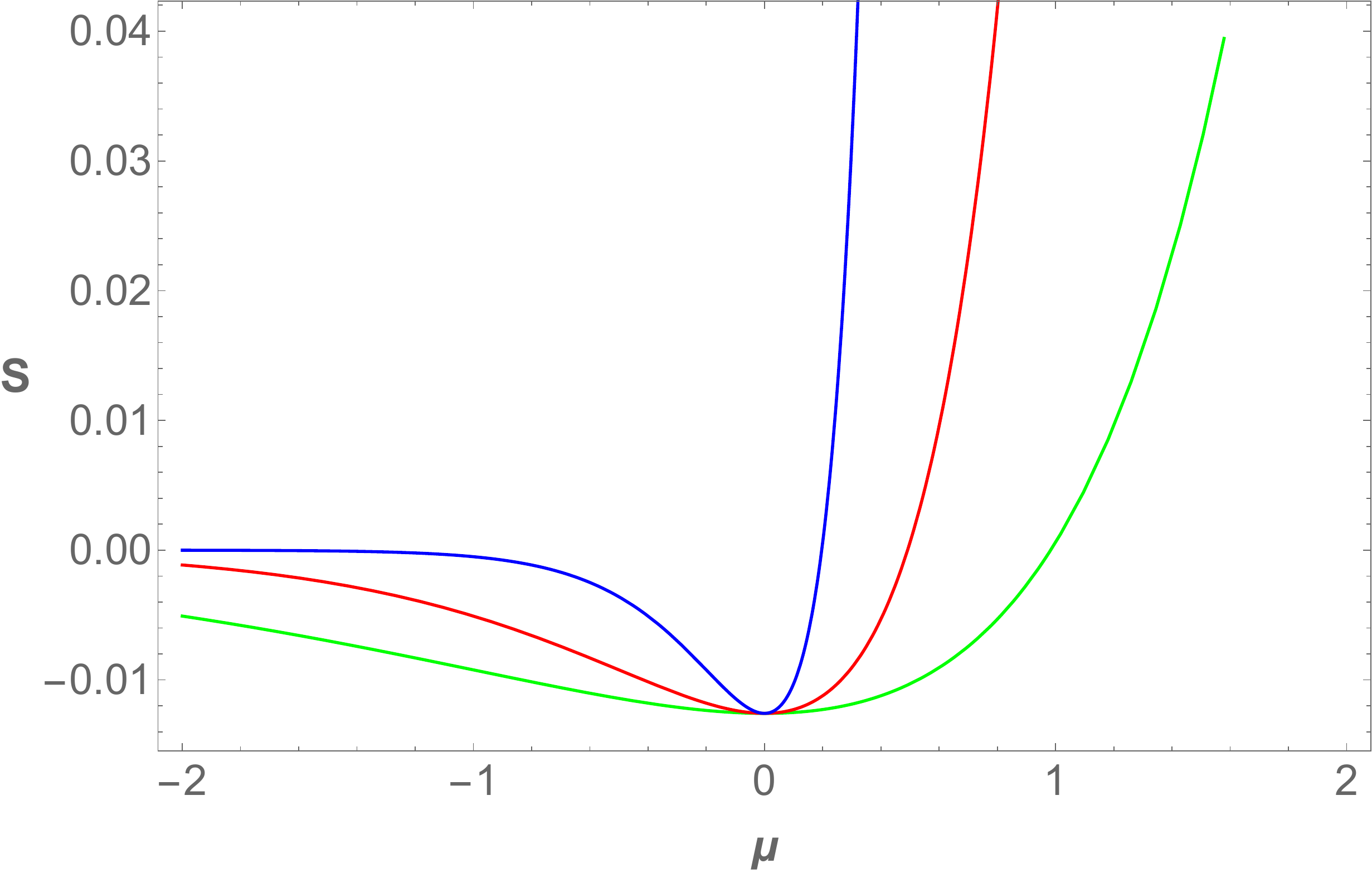}
\includegraphics[width=0.32\textwidth]{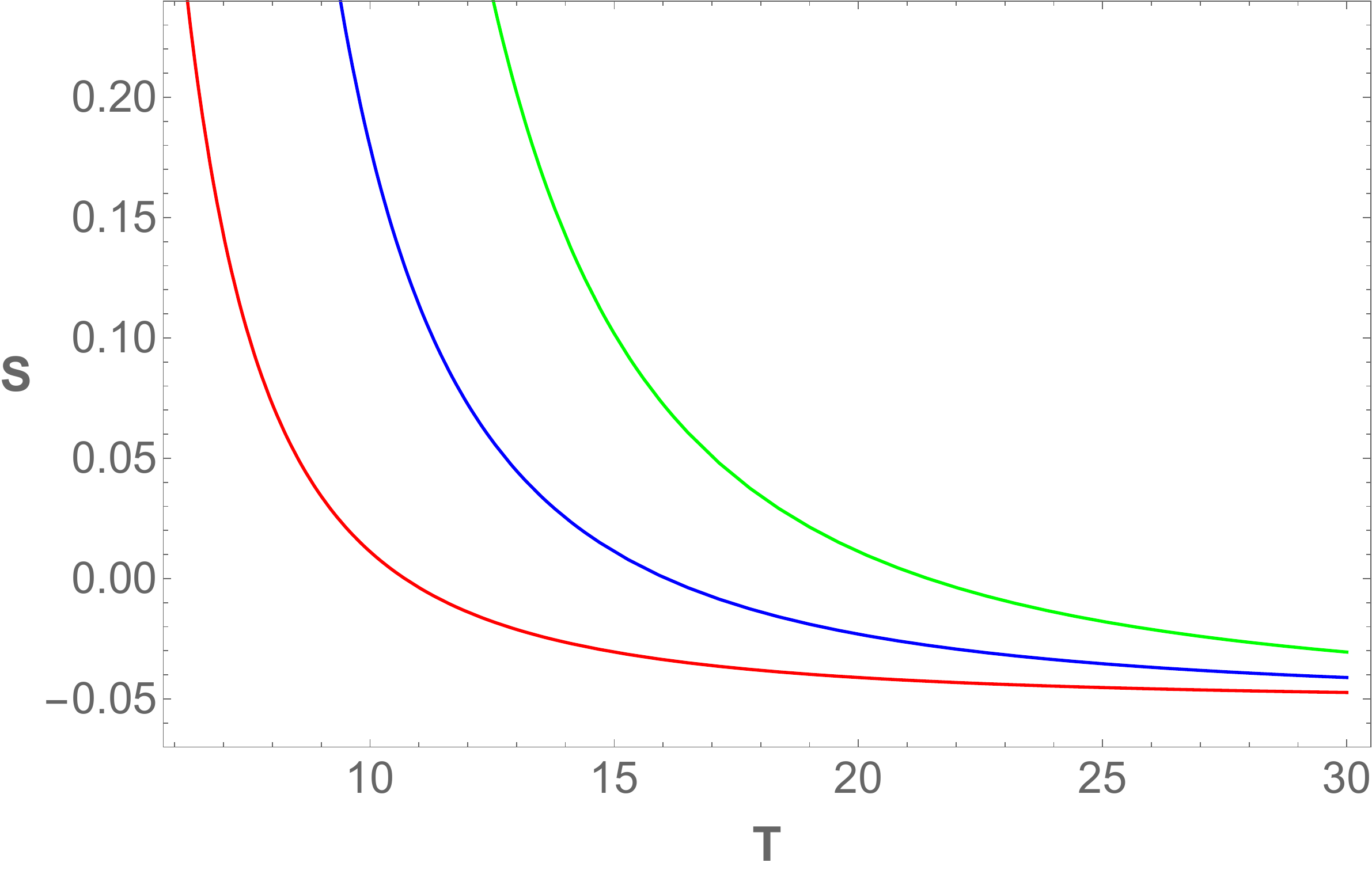}
\includegraphics[width=0.32\textwidth]{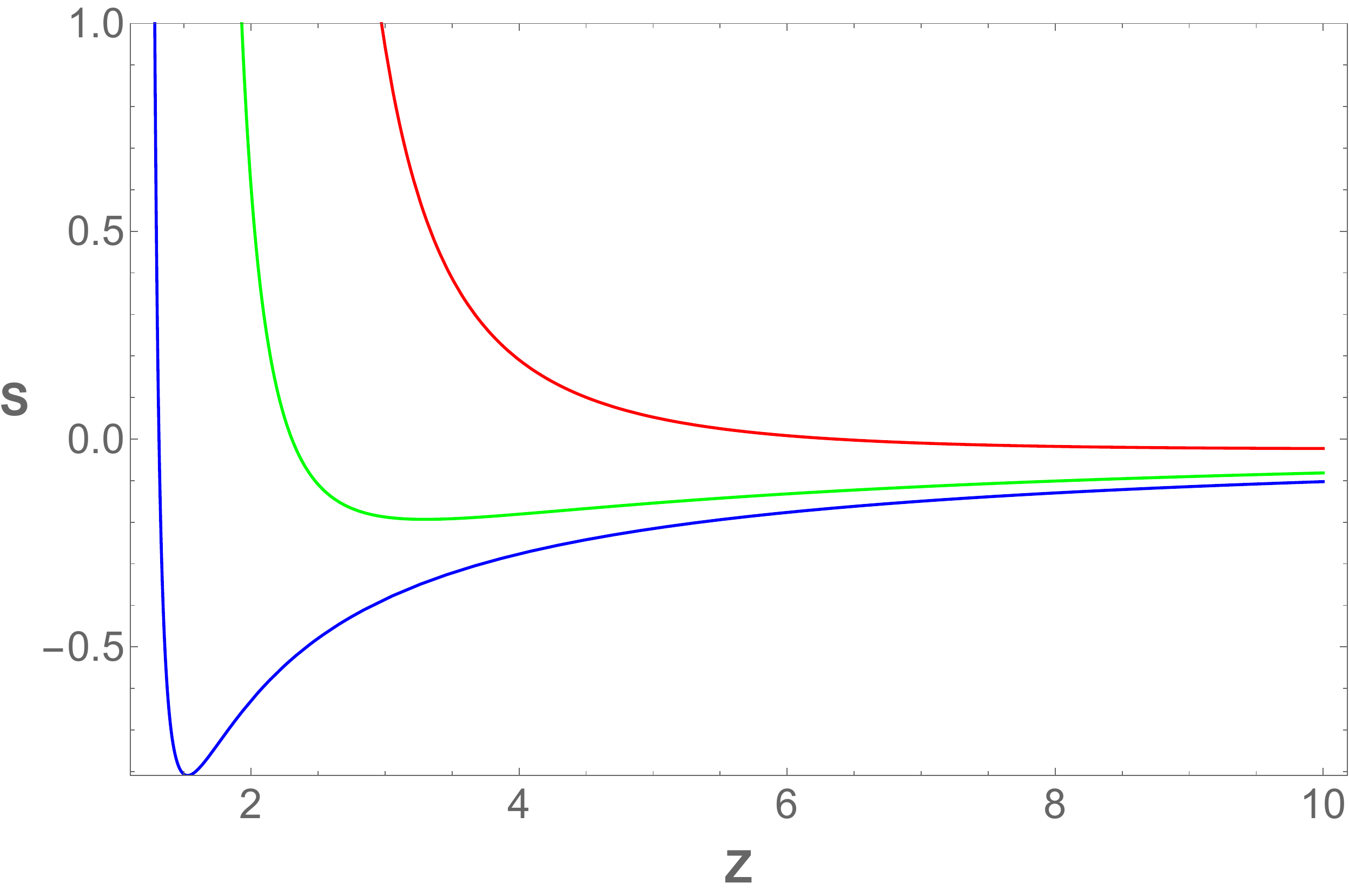}
    \caption{\textit{Left:} Entropy dependence on $\mu$ with: $M=1$ and $Z_1=80$, (a) $T=1$  (green); (b) $T=0.5$ (red); (c) $T=0.2$  (blue). \textit{Middle:} Entropy dependence on $T$ with: $M=1$ and $Z_1=20$, (a) $\mu=20$ (green); (b) $\mu=10$ (red); (c)  $\mu=15$ (blue). \textit{Right:} Entropy dependence  on $Z$ with: $M=1$ and $Z_1=20$, (a) $\mu=20$ (green); (b) $\mu=10$ (red); (c)  $\mu=15$  (blue).}
    \label{fig:Sy}
\end{figure}

In Figure \ref{fig:Sy} the dependence of the entropy $S$  on several parameters is analyzed (for simplicity we took $Z_i(T)$ to be a constant so $\frac{\partial  z_i(T)}{\partial T}=0$) . In Fig.~\ref{fig:Sy}a the entropy's dependence on $\mu$ is plotted. We observe that the entropy has a minimum, and this minimum does not seem to depend on the value of the temperature $T$. Fig.~\ref{fig:Sy}b shows the dependence of the entropy  on $T$. We observe that the entropy is a decreasing function of T, but the rate of this decrease depends on the value of $\mu$. We also investigate the dependence of the entropy as a function of  $Z$ in Fig.~\ref{fig:Sy}c. We see that the entropy has a minimum for $\mu=15$ and $\mu=20$, and for $\mu=10$ it is a decreasing function of $Z$.

By definition the average number of particles $\Bar{N}$ is given by
\begin{equation}
  \Bar{N}=\left(\frac{\partial}{\partial \mu}(T\ln \Xi) \right)_{T,V}
   =\sum_{i=1}^M\frac{e^{\frac{\mu}{T}}}{e^{\frac{\mu}{T}}-z_i(T)}. \\
   \end{equation}
\begin{equation}
\begin{aligned}
  \therefore \Bar{N}=\sum_{i=1}^M\frac{e^{\frac{\mu}{T}}}{e^{\frac{\mu}{T}}-z_i(T)}.
  \label{eqn:nbaryl}
    \end{aligned}
   \end{equation}
By definition, the grand canonical potential is 
\begin{equation}
    \Phi_G=-T\ln \Xi=U-\mu\Bar{N}-TS.
    \label{eqn:gcp}
\end{equation}
where U is the average energy
\begin{equation}
 U=\mu\Bar{N}+TS-T\ln\Xi.
\end{equation}
Substituting Eqs.~(\ref{eqn:nbaryl})-(\ref{eqn:syl})-(\ref{eqn:lnxi}) in the latter equation
\begin{eqnarray}
  U && \resizebox{.94\hsize}{!}{$=\mu \sum_{i=1}^M\frac{e^{\frac{\mu}{T}}}{e^{\frac{\mu}{T}}-z_i(T)}+T \sum_{i=1}^M\left[ln \left(1-\frac{e^{\frac{\mu}{T}}}{z_i(T)}\right)+T\left(\frac{e^{\frac{\mu}{T}}}{z_i(T)-e^{\frac{\mu}{T}}}\right)\left(\frac{\mu}{T^2}+\frac{1}{z_i(T)}\frac{\partial  z_i(T)}{\partial T} \right)\right]$}\nonumber \\
    & & -T\sum_{i=1}^M\ln \left(1-\frac{e^{\frac{\mu}{T}}}{z_i(T)}\right).
     \label{eqn:UY}
\end{eqnarray}

From the definition of the grand canonical potential Eq.~(\ref{eqn:gcp})
  \begin{equation}
   \Phi_G=-T\ln(\Xi)=U-\mu\Bar{N}-TS=-P_V V,
   \end{equation}
from where 
\begin{equation}
    P_V=\frac{T}{V}\ln\Xi=\frac{T}{V}\sum_{i=1}^M\ln \left(1-\frac{z}{z_i(T)}\right) = \frac{T}{V}\sum_{i=1}^M\ln \left(1-\frac{e^{\frac{\mu}{T}}}{z_i(T)}\right).
     \label{eqn:PY}
\end{equation}

  \begin{figure}[t]
    \centering
        \includegraphics[width=0.32\textwidth]{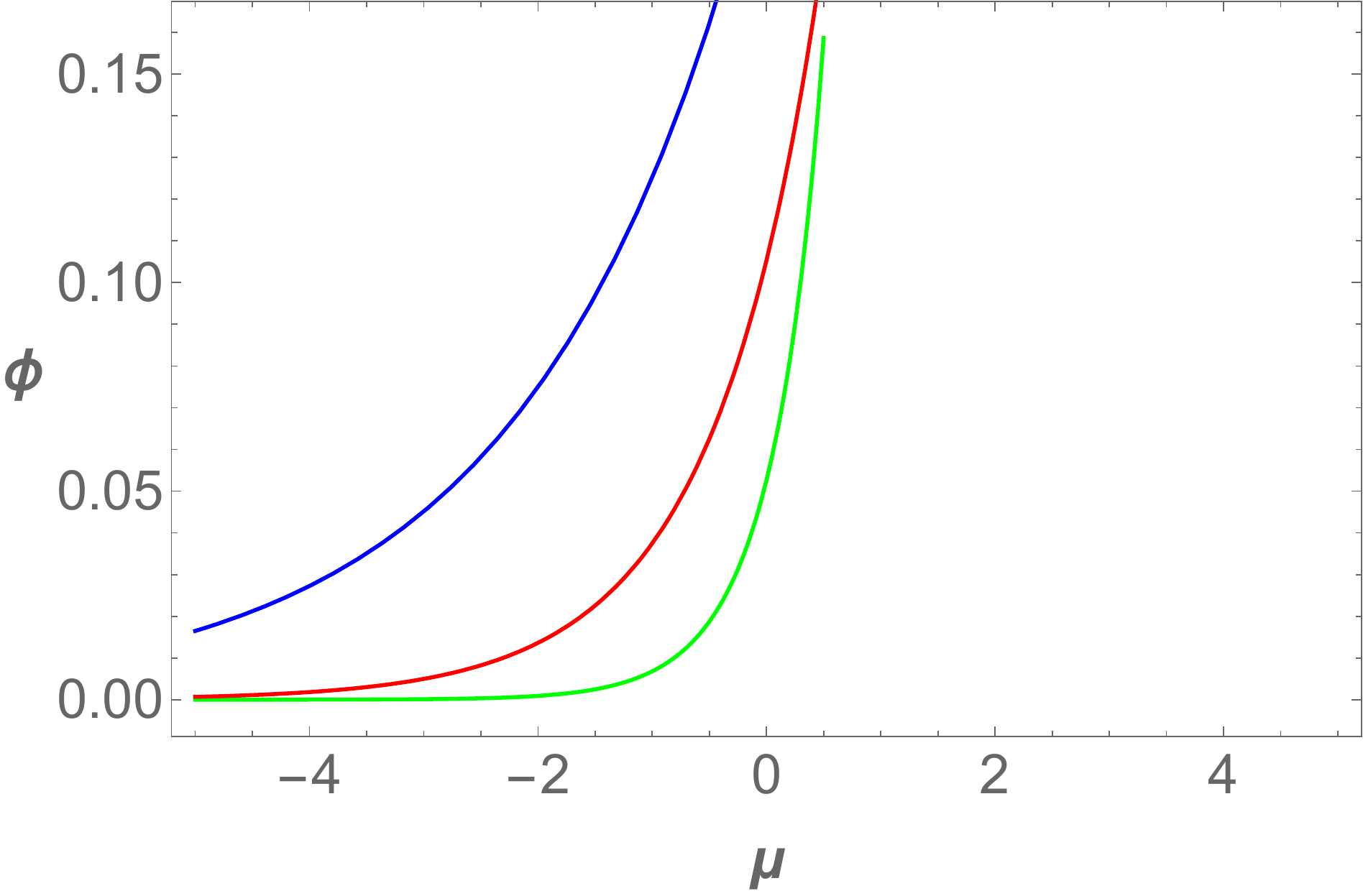}
\includegraphics[width=0.32\textwidth]{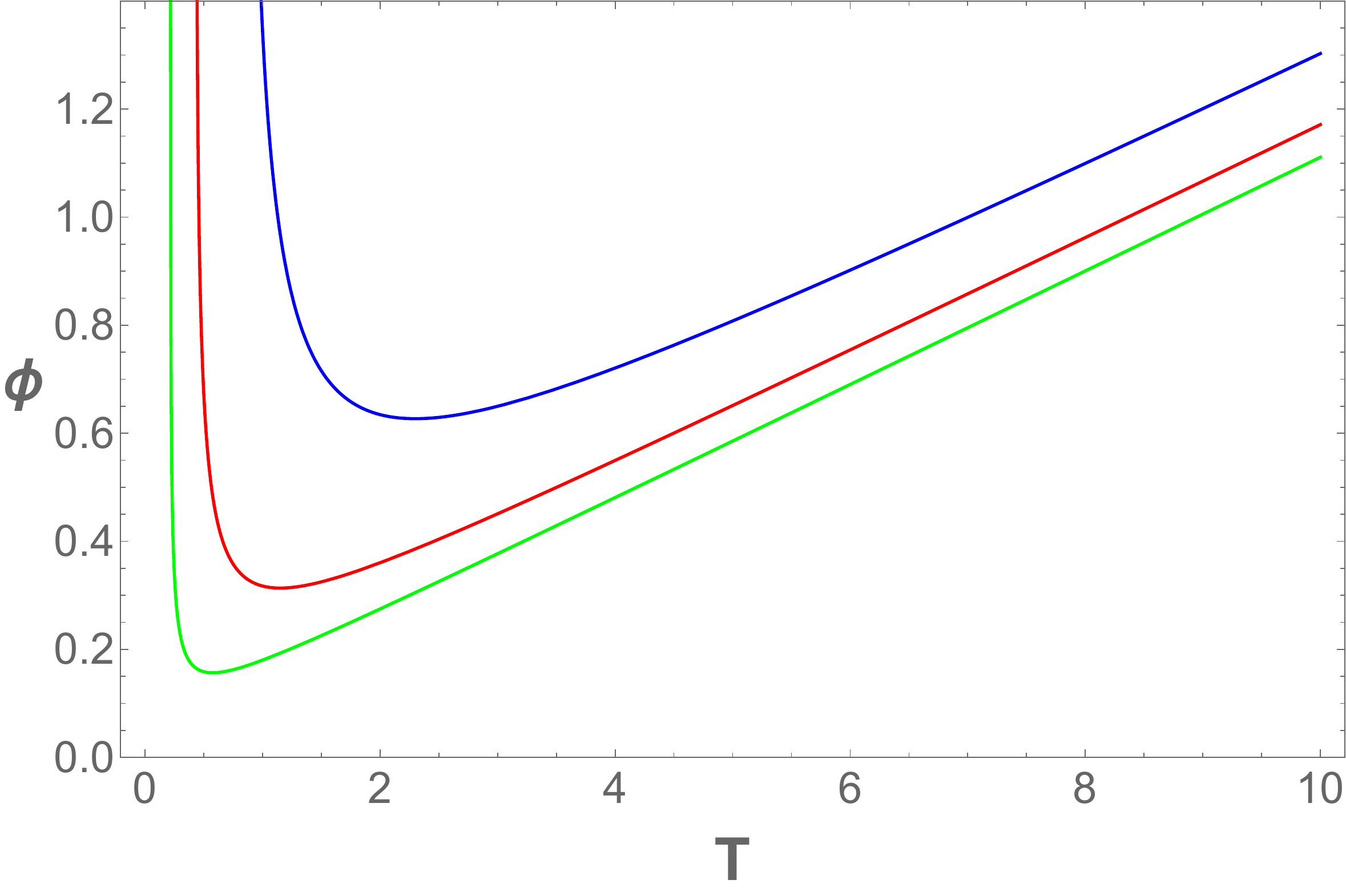}
\includegraphics[width=0.32\textwidth]{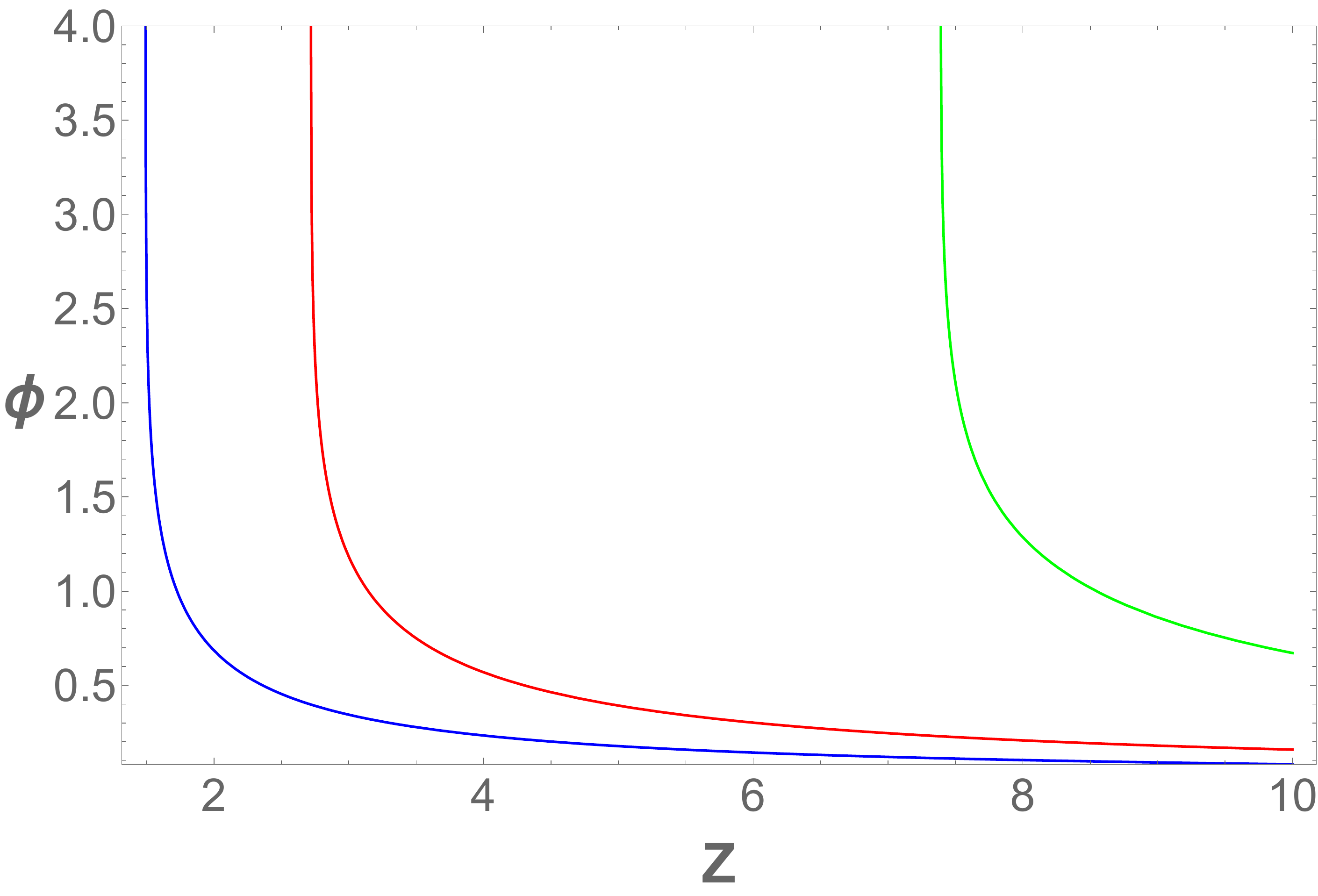}
    \caption{\textit{Left:} Grand canonical potential dependence on $\mu$ with: $M=1$ and $Z_1=10$, (a) $T=0.5$  (green); (b) $T=1$  (red); (c) $T=2$  (blue). \textit{Middle:} Grand canonical potential dependence on $T$ with: $M=1$ and $Z_1=10$, (a) $\mu=0.5$ (green); (b) $\mu=1$ (red); (c)  $\mu=2$  (blue). \textit{Right:} Grand canonical potential dependence on $Z$ with: $M=1$ and $T=0.5$, (a) $\mu=1$ (green); (b) $\mu=0.5$ (red); (c)  $\mu=0.2$  (blue).}
    \label{fig:phig}
\end{figure}

   In  Figure \ref{fig:phig} we analyze the behavior of the grand canonical potential $\Phi_G=\Phi$  (where we changed $\Phi_G$ for $\Phi$ to simplify the notation of the plots) on various parameters. Fig.~\ref{fig:phig}a   presents the grand canonical potential as a function of $\mu$. We see that the grand canonical potential is an increasing function of $\mu$ but the rate of this increase depends on the value of the temperature $T$. In Fig.~\ref{fig:phig}b the grand canonical potential dependence on $T$ is analyzed. We observe that the grand canonical potential has a minimum and this minimum is dependent on the value of $\mu$. Fig.~\ref{fig:phig}c  illustrates the grand canonical potential dependence on $Z$. We can observe that the grand canonical potential is a decreasing function of $Z$, but the rate of this decrease depends on the value of $\mu$

The Gibbs free energy, G, is given by 

\begin{equation}
  G=H-TS=U+P_VV-TS.
  \label{eqn:g}
\end{equation}
Substituting Eqs.~(\ref{eqn:UY})-(\ref{eqn:PY})-(\ref{eqn:syl}) in Eq.~(\ref{eqn:g}).
\begin{equation}
     G     =\mu \sum_{i=1}^M\frac{e^{\frac{\mu}{T}}}{e^{\frac{\mu}{T}}-z_i(T)}.
\end{equation}

Defining $\lambda(T,\mu)=\frac{z_i(T)}{e^{\frac{\mu}{T}}}$
\begin{equation}
\begin{aligned}
     G=\mu \sum_{i=1}^M\frac{1}{1-\frac{z_i(T)}{e^{\frac{\mu}{T}}}}=\mu \sum_{i=1}^M\frac{1}{1-\lambda(T,\mu)}.
\end{aligned}
\end{equation}

\begin{figure}[t]
    \centering
         \includegraphics[width=0.32\textwidth]{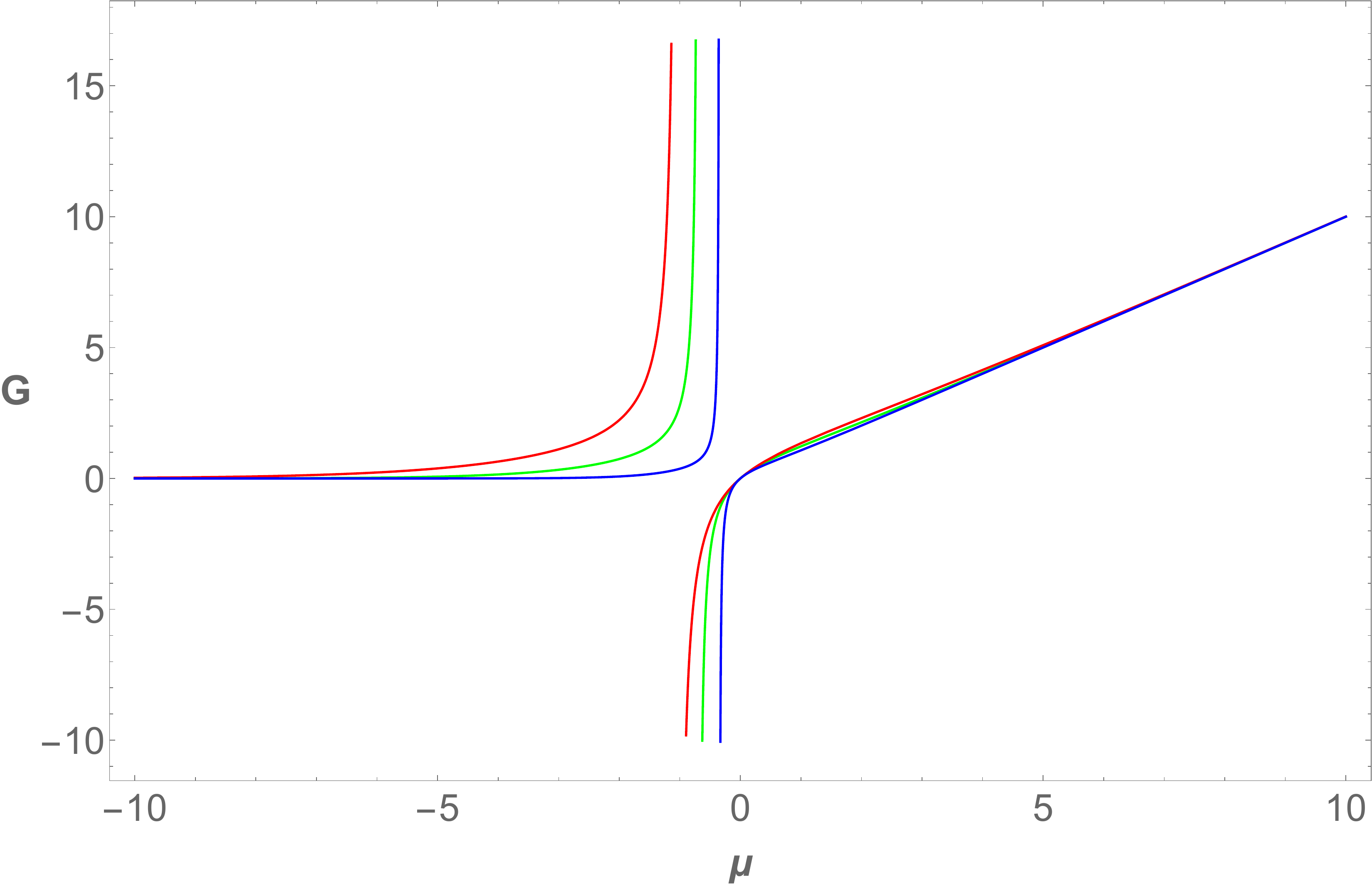}
\includegraphics[width=0.32\textwidth]{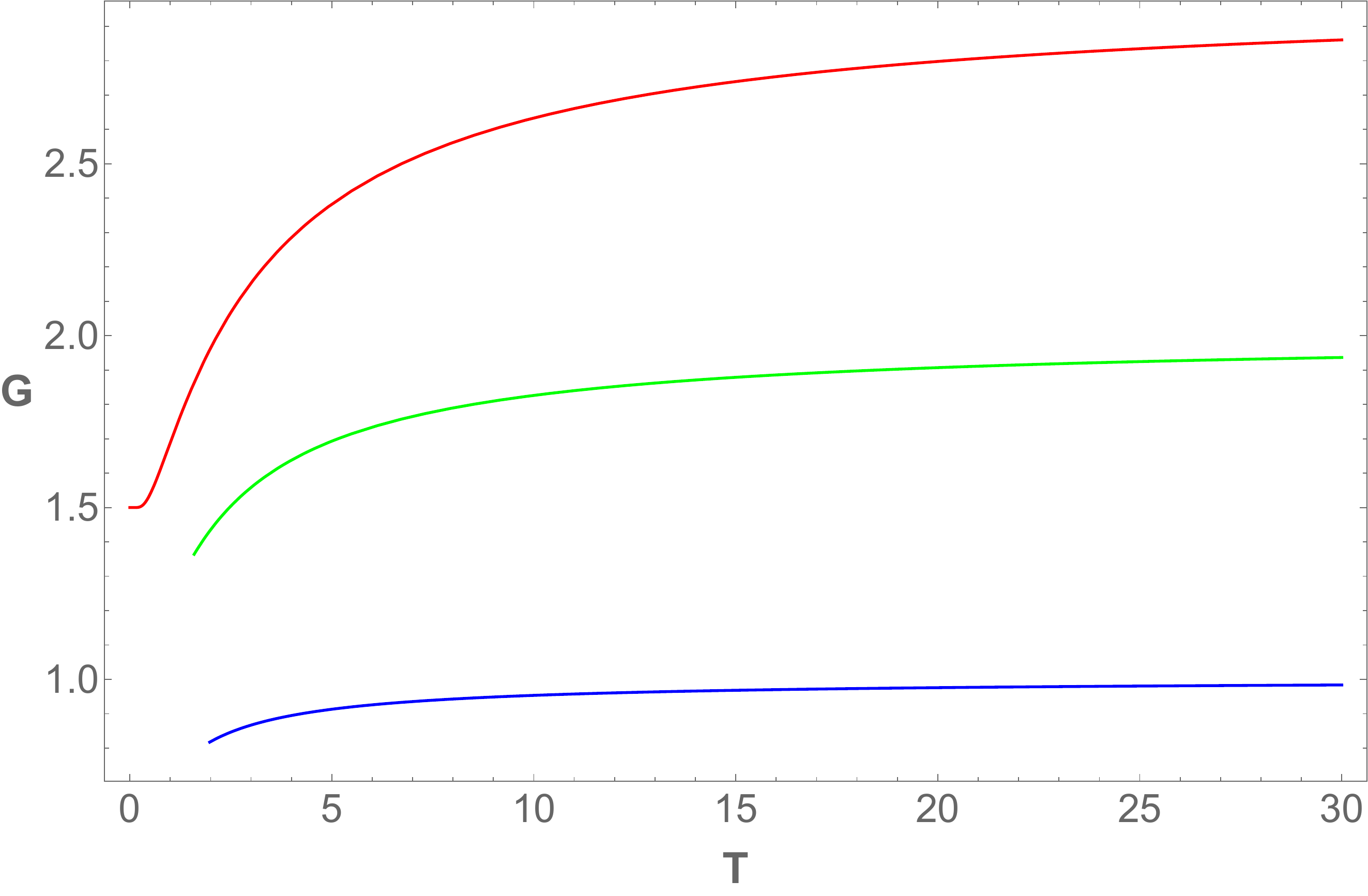}
\includegraphics[width=0.32\textwidth]{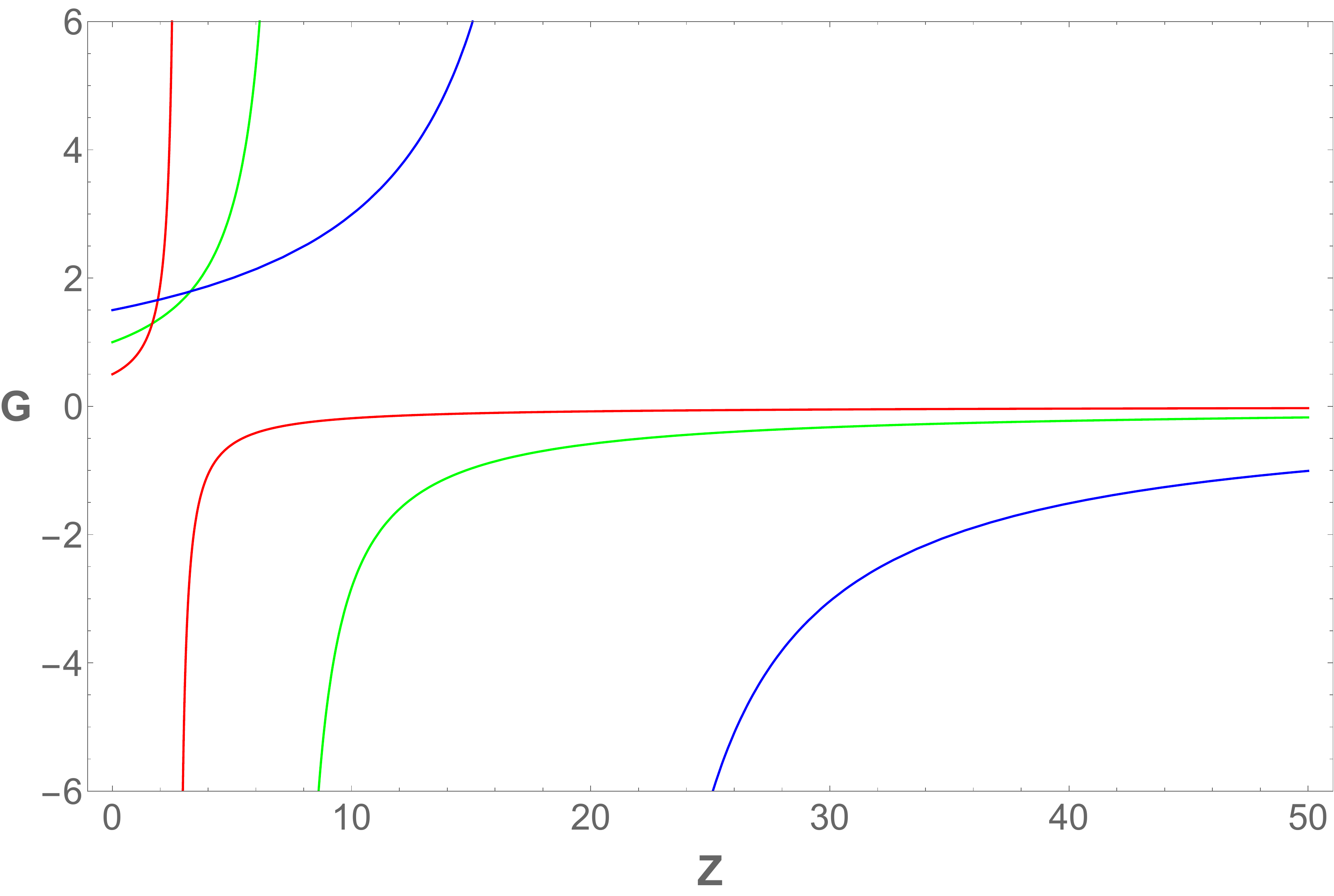}
    \caption{\textit{Left:} Gibbs free energy dependence on $\mu$ with: $M=1$ and $Z=0.5$, (a) $T=1$  (green); (b) $T=1.5$ (red); (c) $T=0.5$  (blue). \textit{Middle:} Gibbs free energy dependence on $T$ with:  $M=1$ and $Z=0.5$, (a) $\mu=1$ (green); (b) $\mu=1.5$ (red); (c)  $\mu=0.5$  (blue). \textit{Right:} Gibbs free energy dependence on $Z$ with: $M=1$ and $T=0.5
    $, (a) $\mu=1$ (green); (b) $\mu=0.5$ (red); (c)  $\mu=1.5$  (blue).}
    \label{fig:gfe}
\end{figure}

 In  Figure \ref{fig:gfe} we analyze the dependence of the Gibbs free energy $G$ on several parameters. Fig.~\ref{fig:gfe}a show the Gibbs free energy behavior on $\mu$. We see that the Gibbs free energy has a discontinuity but it appears to change with the value of the temperature $T$. Fig.~\ref{fig:gfe}b  illustrates the Gibbs free energy as a function of $T$. We observe that the Gibbs free energy is an increasing function of $T$ but the rate of this increase depends on the value of $\mu$. Fig.~\ref{fig:gfe}c  exhibits the dependence of the  Gibbs free energy dependence on $Z$. We see that there is a discontinuity of the Gibbs free energy, and this discontinuity seems to depend on the value of $\mu$.

\section{Yang-Lee theory and its Gravitational Phase Transition}
 \label{sec:Yang-Lee}

The presence of a phase transition has been found to be related to the vanishing of the grand canonical partition function in the fugacity plane by Yang and Lee\cite{YL}. Consequently, we search for the zeros of the fugacity  (\emph{cf}. Saslaw \emph{et al}. \cite{Saslaw}). 

 From $z=e^{\frac{\mu}{T}}$ and from Eq.~(\ref{eqn:mu}) we explicitly calculate the activity: \\
\begin{equation}
    z=\rho \left(\frac{2\pi MT}{\lambda^2} \right)^{-\frac{3}{2}}(1+\alpha x)^{-1}e^{-B_l}.
    \label{eqn:zactivity}
\end{equation}
If $z$ is considered to be complex and we also consider the thermodynamic limit in which both $N \rightarrow \infty$ and the volume of the system $V \rightarrow \infty$, such that the density $N/V \rightarrow \rho$ is constant. Then, we see that the activity $z$ is  positive in the thermodynamic limit for any value of N. For a given thermodynamic limit $\rho$, $z$ will be zero if $x\rightarrow \infty$ or $B_l \rightarrow \infty$. From  Eq.~(\ref{eqn:Bl})  we see that $0 \leq B_l \leq 1 $, so $B_l= \infty$ is not a valid condition. Also, from Eq.~(\ref{eqn:x}) if $x\rightarrow \infty$ then $T \rightarrow 0$. But zero temperature is unwanted even as an initial condition.

Now we can consider $\rho > 0$ and $T \rightarrow \infty$ which makes the activity $z \rightarrow 0$. 
Consequently, the only possible phase transition, as defined in Yang–Lee theory, for a logarithmic correction to the Newtonian (interaction) potential energy takes place at $x=0$ (which implies that $B_l=0$). This initial condition initiates a hierarchical phase transition that if it were able to occur simultaneously on all scales, would resemble a first-order phase transition. \emph{(d.h. Umwandlung erster Ordnung}).
The system goes through some states that can be thought as second-order phase transitions (\emph{d.h. Umwandlungen
zweiter Ordnung}) and are called milestones.
\begin{itemize}
\item The first milestone happens at $B_l=\frac{1}{3}$ when the specific heat $C_V$ reaches its maximum value, \emph{i.e.}, $C_V=\frac{5}{2}$. The temperature at this point is $
    T=(2\alpha\beta \rho)^{\frac{1}{3}}.$
 \item The second milestone happens at $B_l\approx 0.87$ when the  specific heat $C_V$ becomes negative. The temperature at this point is
$
T=(0.15 \alpha\beta \rho)^{\frac{1}{3}}.
$
\item The third milestone happens at $B_l=1$ when $C_V=-\frac{3}{2}$. The temperature at this point is $T=0.$
\end{itemize}

\begin{figure}[t]
    \centering
    \includegraphics[width=0.5\textwidth]{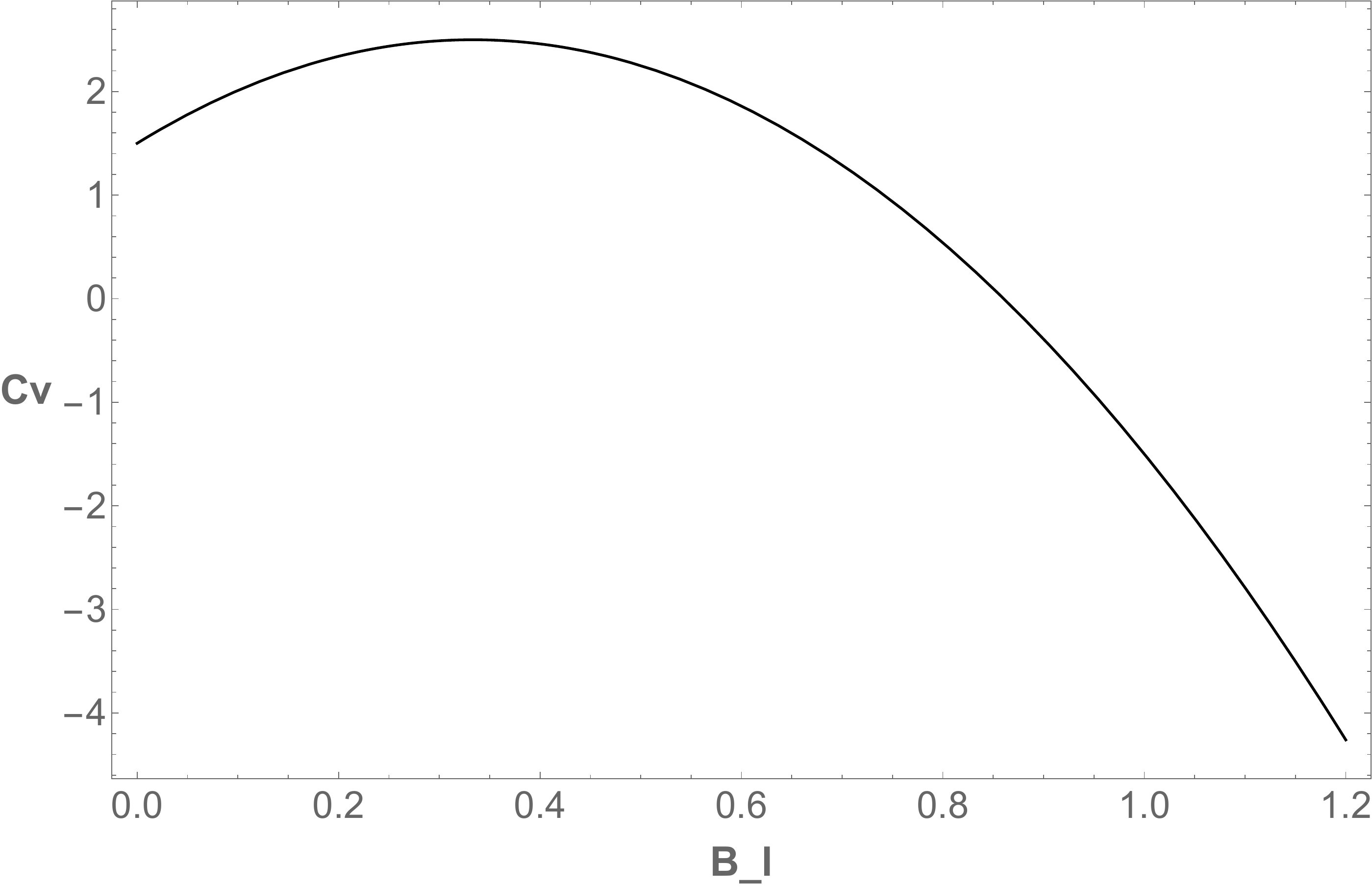}
    \caption{Evolution of $C_V$  Eq.~(\ref{eqn:cv1}) with $B_l$.} 
        \label{fig:Cvbl}
\end{figure}

\section{Specific heat and the effective potential}
  \label{sec:Specific Heat}
  
A physical consequence of an indicator of a hierarchical phase transition is the variation of the specific heat, from  Eq.~(\ref{eqn:ieu}) we have:

 \begin{equation}
    C_V=\frac{1}{N}\left(\frac{\partial U}{\partial T}\right)=\frac{3}{2}(1+4B_l-6B_l^2).
    \label{eqn:cv1}
\end{equation}

\begin{figure}[t]
    \centering
    \includegraphics[width=0.5\textwidth]{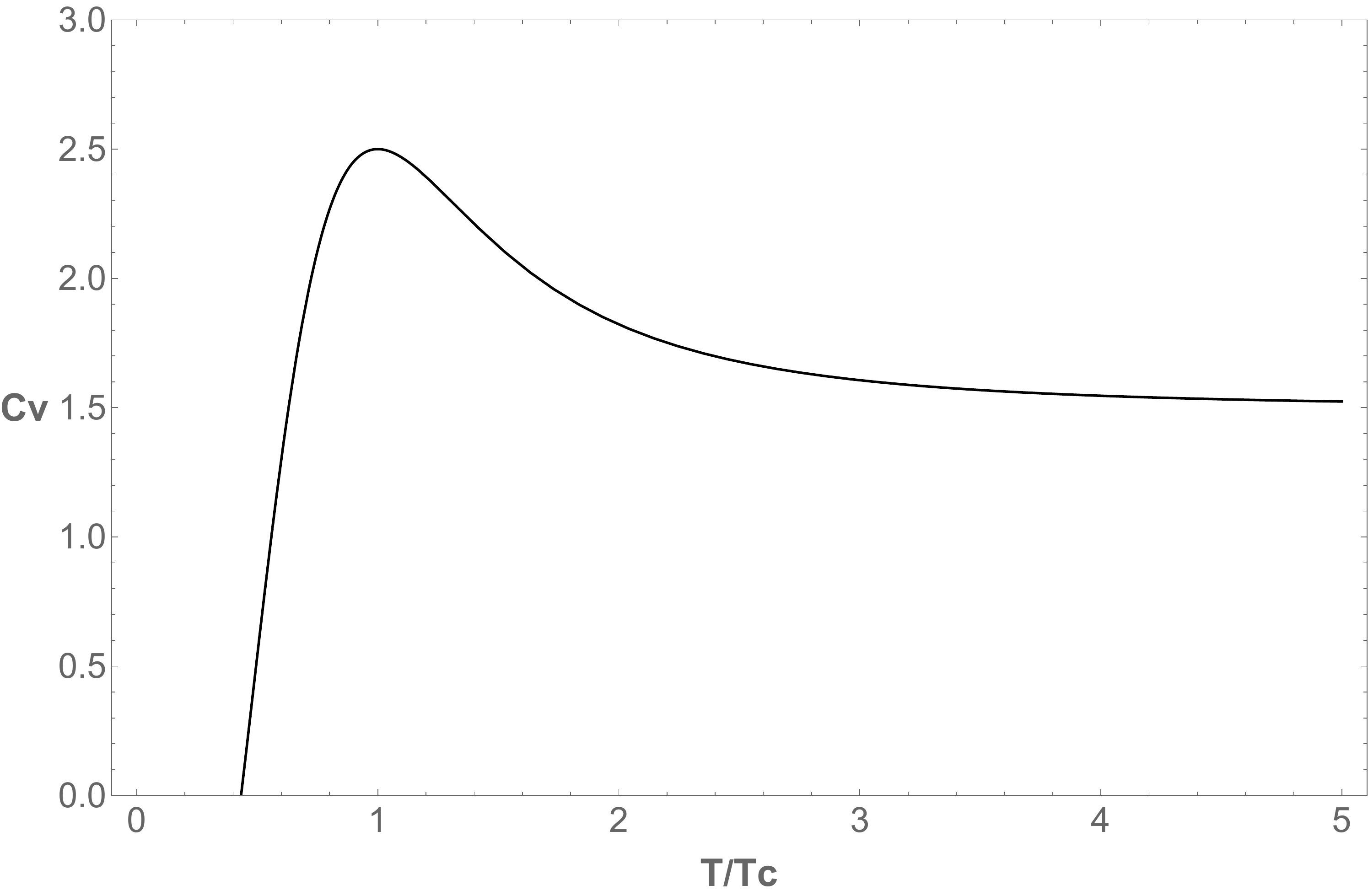}
     \caption{Evolution of $C_V$  Eq.~(\ref{eqn:cv2}) with $\frac{T}{T_c}$. 
     }
    \label{fig:cvtc}
\end{figure}

In Figure \ref{fig:Cvbl}, we analyze the dependence of the specific heat on $B_l$. We see that when $B_l= 0$ the specific heat is  $C_V = \frac{3}{2}$, but when  $B_l= 1$ the specific heat is $C_V = - \frac{3}{2}$. Thus, in betweeen these values the specific heat $C_V$ is a maximum at a critical value for temperature $T=T_C$ given by 
\begin{equation}
    \frac{\partial C_V}{\partial T}=0.
\end{equation}
From which we obtain the critical temperature $T_C$
\begin{equation}
   T_C=(2\alpha \beta \rho)^{\frac{1}{3}}.
\end{equation}
Writing the specific heat $C_V$ in terms of $T_c$ 
\begin{equation}
    C_V=\frac{3}{2}\left[]1-2\left(\frac{1-4\left(\frac{T}{T_c} \right)^3}{(1+2\left(\frac{T}{T_c} \right)^3)^2}\right) \right].
    \label{eqn:cv2}
\end{equation}
In Figure \ref{fig:cvtc}, we analyze the dependence of the specific heat on $\frac{T}{T_c}$. We see that the specific heat reaches its maximum value $C_V=\frac{5}{2}$ at  $\frac{T}{T_c}=1$.

  The effective potential of a scalar field distribution surrounding a black hole obtained from exploring the analogy between the Klein–Gordon equation and the Gross–Pitaevskii equation in a Schwarzschild spacetime background was deduced by Castellanos \emph{et al} \cite{Cast}.

The effective trapping potential is
\begin{equation}
    V_{\text{eff}_S}=F\left(\sigma^2+\frac{F'}{r^*}\right),
\end{equation}
where the prime notation means the derivative of F concerning  $r^*=\int\frac{dr}{F}$.

\begin{figure}[t]
    \centering
    \includegraphics[width=0.5\textwidth]{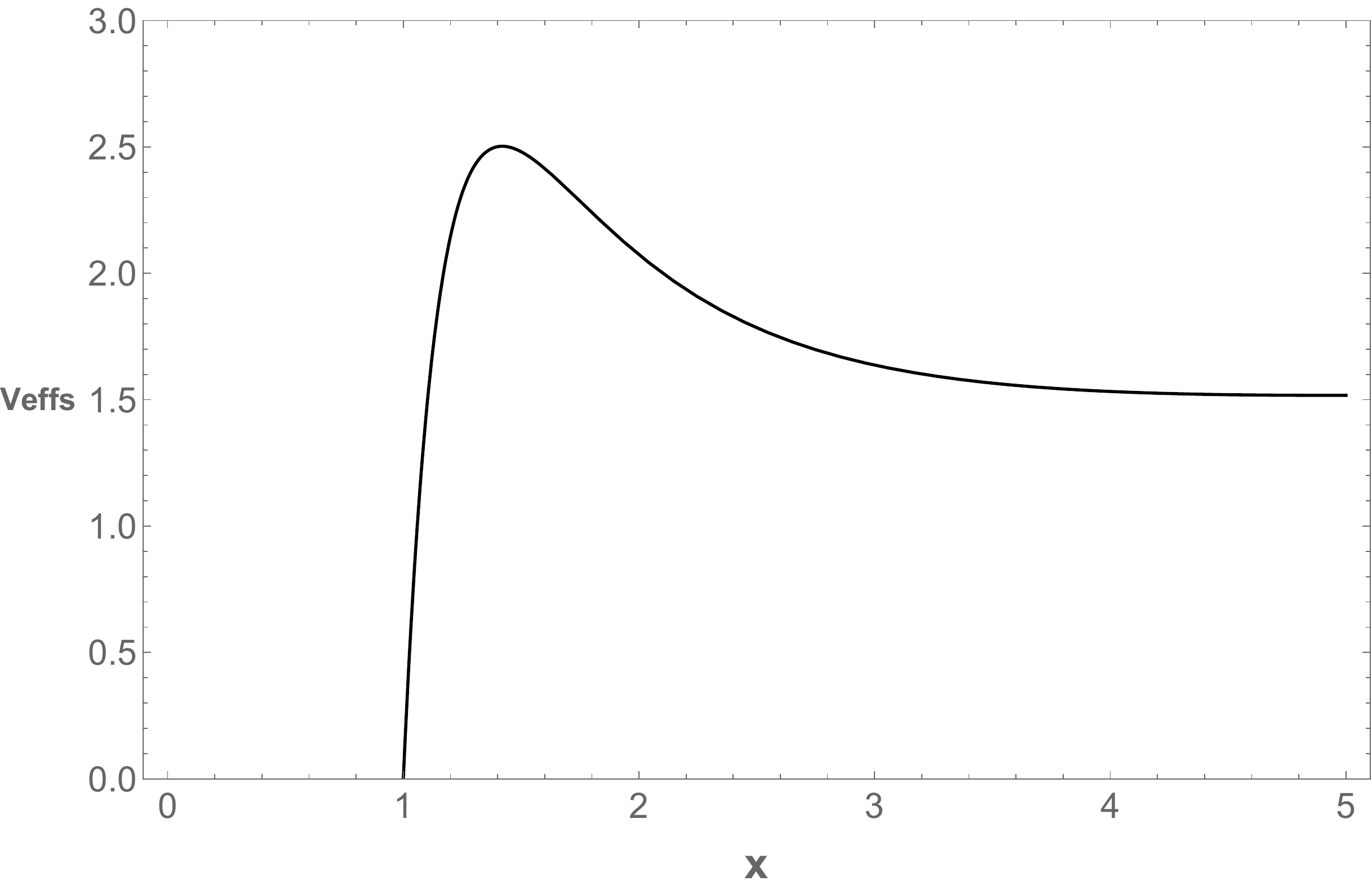}
     \caption{Effective potential  Eq.~(\ref{eqn:veff}), where we have consider $R_0=0.228$, $q=1$, $n=0.5$, $\alpha=0.301$. 
     }
    \label{fig:ve}
\end{figure}

Considering a Schwarzschild spacetime configuration we have
\begin{equation}
       F=1-\frac{2MG}{c^2r}=1-2q\frac{n}{x},
   \end{equation}
from where we get the effective potential
\begin{equation}
    V_{\text{eff}_S}=\frac{1}{R_0^2}\left(\alpha^2+\frac{2qn}{x^3} \right)\left( 1- \frac{2qn}{x}\right).
    \label{eqn:veff}
\end{equation}
In Figure \ref{fig:ve}, we analyze the dependence of the effective potential on $x$. For the specific configuration $R_0=0.228$, $q=1$, $n=0.5$ and $\alpha=0.301$  at $x\approx1.419$ the effective potential is $V_{\text{eff}_S}\approx 2.50262$ and it is a maximum value. Comparing Fig.~\ref{fig:Cvbl} and Fig.~\ref{fig:ve} we consider the specific heat $C_V$ Eq.~(\ref{eqn:cv2}) (as function of $\frac{T}{T_c}$) as a effective potential $V_{eff_S}$ Eq.~(\ref{eqn:veff}) with the parameters $R_0=0.228$, $q=1$, $n=0.5$ and $\alpha=0.301$. 

\section{Discussion and Conclusions}
  \label{sec:conclusions}

In this paper, we have analyzed the effects of a nonlocal gravitational potential correction to the Newtonian (interaction) potential energy in the thermodynamic quantities of a large system of galaxies, where each galaxy was treated as a point particle in a grand canonical ensemble through the gravitational partition function. We have explored the possibility of a phase transition using Yang-Lee theory  where some types of phase transitions can be described by the zeros of the grand canonical partition function in the fugacity plane. Given that we were able to find the activity explicitly it was possible to determine the real zeros. We notice that none of them falls between $B_l=0$ and $B_l=1$, although we saw that the specific heat changes sign, in the sense of Yang-Lee theory, we have no phase transition. Nonetheless, for $x=0$ ($B_l=0$) which corresponds to a system with a temperature that tends arbitrarily near to infinity, the fugacity tends arbitrarily close to zero. This is what we called \textit{the start of a phase transition}. 

Our results present some modifications of the work done by Capozziello \emph{et al} \cite{Capo}, where we varied some parameters in the thermodynamic equations and see how the evolution of the different thermodynamic quantities indeed changed. In particular, as shown in Fig.~\ref{fig:Helmholtz}a the Helmholtz free energy is still a decreasing function of $x$ for a large value of the nonlocality parameter $\lambda$, but for Fig.~\ref{fig:Helmholtz}b and Fig.~\ref{fig:Helmholtz}c the plots differ from the ones provided by Capozziello \emph{et al}\cite{Capo}. Analogously, we obtained the same tendency of the Fig.~\ref{fig:Interenergy}a of the internal energy $U$ to be a decreasing function of $x$ for a large value of $\lambda$ and in Fig.~\ref{fig:Interenergy}c the internal energy has a minimum for $\sigma$ that does not seem to depend on the temperature $T$ of the system. The distribution function $F(N)$ in Eq.~(\ref{eqn:dfn}) is the probability for finding a cell with precisely $N=\Bar{N}$ particles  within cells in a grand canonical ensemble that in a volume $V$ have an average number of particles $\Bar{N}$. Therefore, the behavior in Fig.~\ref{fig:F(N)} shows the evolution of the distribution function for the specific case $N=\Bar{N}$ concerning the different parameters considered. We also abridge from a standard statistical mechanics frame (with the partition function)  the entropy $S$ and from a modified statistical mechanics frame (considering a modification to the grand canonical partition function from its zeros) the entropy $S$, the grand canonical potential $\Phi$, and the Gibbs free energy $G$.
The conditions found for the phase transition of the system with a logarithmic correction to the Newtonian (interaction) potential energy resemble the ones found by Saslaw \emph{et al} \cite{Saslaw} for considering only a Newtonian (interaction) potential energy between particles. 
We found a remarkable connection between the specific heat $C_V$ Eq.~(\ref{eqn:cv2}) and the effective potential $V_{\text{eff}_S}$ Eq.~(\ref{eqn:veff}) for the specific parameters $R_0=0.228$, $q=1$, $n=0.5$ and $\alpha=0.301$, an extension of this work will be the study of a deeper connection between them, which will be reported elsewhere. 

Our finding highlights the possibility of studying the phase transition of a system with a nonlocal correction of gravity through the analysis of the zeros of the nonlocal gravitational partition function in the fugacity plane.

\begin{acknowledgments}
CE-R acknowledges the Royal Astronomical Society as FRAS 10147. CE-R and CAA are supported by PAPIIT UNAM Projects IA100220 and TA100122.
This work is part of the Cosmostatistics National Group (\href{https://www.nucleares.unam.mx/CosmoNag/index.html}{CosmoNag}) project.

\end{acknowledgments}

\appendix

\section{Extra calculations for the Partition Function}
\label{app:def}

In this appendix we show the hypergeometric function derivation in Eq.~(\ref{eqn:hyper}).

By definition
\begin{equation}
    {}_2F_1\left(a,b;c;x\right)=\frac{\Gamma(c)}{\Gamma(a)\Gamma(b)}\sum_{n=0}^\infty\frac{\Gamma(a+n)\Gamma(b+n)}{\Gamma(c+n)}\frac{x^n}{n!}.
\end{equation}
 If b=c 
\begin{equation}
    \begin{aligned}
    {}_2F_1\left(a,b;b;x\right)&=\frac{\Gamma(b)}{\Gamma(a)\Gamma(b)}\sum_{n=0}^\infty\frac{\Gamma(a+n)\Gamma(b+n)}{\Gamma(b+n)}\frac{x^n}{n!}=\frac{1}{\Gamma(a)}\sum_{n=0}^\infty\Gamma(a+n)\frac{x^n}{n!}\\
    &=\sum_{n=0}^\infty \frac{\Gamma(a+n)}{\Gamma(a)}\frac{x^n}{n!},\\
    \end{aligned}
\end{equation} 
 We see that
 \begin{equation}
    \Gamma(a+n)=a(a+1)(a+2)\ldots(a+n)\Gamma(a),
 \end{equation}
 \begin{equation}
   \Rightarrow \frac{\Gamma(a+n)}{\Gamma(a)}=a(a+1)(a+2)\ldots(a+n),
 \end{equation}
\begin{equation}
    \begin{aligned}
    {}_2F_1\left(a,b;b;x\right)
    &=\sum_{n=0}^\infty a(a+1)(a+2)\ldots(a+n)\frac{x^n}{n!}\\ 
    &=1+ax+\frac{a(a+1)x^2}{2!}+\frac{a(a+1)(a+2)x^3}{3!} \ldots
    \label{eqn:2f1}
    \end{aligned}
\end{equation}  
 From the Binomial theorem we see that
 \begin{equation}
 (\alpha+\beta)^n=\alpha^n+n\alpha^{n-1}\beta+\frac{n(n-1)}{2!}\alpha^{n-2}\beta^2 +\ldots
 \end{equation}
 If $\alpha=1 ,\beta=-x$ and  $n=-a$
 \begin{equation}
 \begin{aligned}
 \Rightarrow (1-x)^{-a} &=1^{-a}+(-a)(1)^{-a-1}(-x)+(1)^{-a-2}\frac{-a(-a-1)(-x)^2}{2!}\\
 &=1+ax+\frac{a(a+1)x^2}{2!} +\frac{a(a+1)(a+2)x^3}{3!}\ldots 
 \label{eqn:1-x}
  \end{aligned}
 \end{equation} 
Comparing  \ref{eqn:1-x} and  \ref{eqn:2f1}
\begin{equation}
    \begin{aligned}
    {}_2F_1\left(a,b;b;x\right)
    &=(1-x)^{-a},
    \end{aligned}
\end{equation}   
 
\begin{equation}
 \begin{aligned}
  \Rightarrow    \frac{1}{(\left(\frac{r_{12}}{\epsilon}\right)^2+ 1)^{\frac{n+l}{2}}} &=
         {}_2F_1\left(\frac{n+l}{2},b;b;-\left(\frac{r_{12}}{\epsilon}\right)^2\right).
         \end{aligned}
\end{equation}


\bibliographystyle{ieeetr}
\bibliography{sample.bib}


\end{document}